\documentclass[fleqn,usenatbib]{mnras}

\usepackage[T1]{fontenc}
\usepackage{ae,aecompl}

\usepackage{graphicx}    
\usepackage{amsmath}    
\usepackage{amssymb}    
\usepackage{stfloats}
\usepackage{pdflscape} 

\newcommand{\Ha}{H$\alpha$}
\newcommand{\Hb}{H$\beta$}
\newcommand{\rNii}{$[\mathrm{N}\,{\textsc{ii}}]\,\lambda6585$}
\newcommand{\bNii}{$[\mathrm{N}\,{\textsc{ii}}]\,\lambda6550$}
\newcommand{\Nii}{$[\mathrm{N}\,{\textsc{ii}}]$}
\newcommand{\rOiii}{$[\mathrm{O}\,{\textsc{iii}}]\,\lambda5008$}

\newcommand{\Oiii}{$[\mathrm{O}\,{\textsc{iii}}]$}
\newcommand{\OH}{$12+\log \left ( \mathrm{O/H} \right )$}

\title[Search for XMP with CNNs]{Efficient Search for Extremely Metal Poor Galaxies in the Local Universe using Convolutional Neural Networks} 
\author[Ting-Yun Cheng et al.]{Ting-Yun~Cheng,$^{1}$\thanks{E-mail:ting-yun.cheng@durham.ac.uk}
Ryan~J.~Cooke$^{1}$
\\ \\
$^{1}$ Centre for Extragalactic Astronomy, Department of Physics, Durham University, South Road, Durham DH1 3LE, UK\\
}

\date{Accepted XXX. Received YYY; in original form ZZZ}

\pubyear{2024}

\begin{document}
\label{firstpage}
\pagerange{\pageref{firstpage}--\pageref{lastpage}}
\maketitle

\begin{abstract}
Nearby extremely metal-poor galaxies (XMPs) allow us to study primitive galaxy formation and evolution in greater detail than is possible at high redshift. This work, for the first time, promotes the use of convolutional neural networks (CNNs) to efficiently search for XMPs in multi-band imaging data based on their predicted N2 index (N2\,$\equiv\log$\{\rNii/\Ha\}). We developed a sequential characterisation pipeline, composed of three CNN procedures: (i) a classifier for metal-poor galaxies, (ii) a classifier for XMPs, and (iii) an N2 predictor. The pipeline is applied to over 7.7 million SDSS DR17 imaging data without SDSS spectroscopy. The predicted N2 values are used to select promising candidates for observations. This approach was validated by new observations of 45 candidates with redshifts less than 0.065 using the 2.54~m Isaac Newton Telescope (INT) and the 4.1~m Southern Astrophysical Research (SOAR) Telescope between 2023 and 2024. All 45 candidates are confirmed to be metal-poor, including 28 new discoveries. 
There are 18/45 galaxies lacking detectable \rNii\ lines ($S/N<2$); for these, we report $2\sigma$ upper limits on their oxygen abundance. Our XMPs have estimated oxygen abundances of $7.1\leq$\OH$\leq8.7$ ($2\sigma$ upper limit), based on the N2 index, and 21 of them with estimated metallicity $<0.1~Z_{\odot}$. 
Additionally, we identified 4 potential candidates of low-metallicity AGNs at $\lesssim0.1Z_{\odot}$. Finally, we found that our observed samples are mostly brighter in the $g-$band compared to other filters, similar to blueberry (BB) galaxies, resembling green pea galaxies and high-redshift Ly$\alpha$ emitters.


\end{abstract}


\section{Introduction}
\label{sec:intro}
Extremely metal-poor galaxies (hereafter XMP) are commonly defined to have a gas-phase metallicity ten times lower than the Sun \citep{Kunth2000}. Some of the well-known examples include I Zwicky 18 \citep{Sargent1970} and SBS 0335-052 \citep{Izotov1997}, each having a metallicity of $\simeq1/30$ $Z_\odot$. Due to selection effects, the XMPs are mostly star-forming dwarf galaxies such as blue compact dwarf galaxies (BCD), that are characterized by prominent hydrogen emission lines. They tend to be less massive ($10^6-10^8M_{\odot}$) but contain massive stars, harbour near-pristine gas, and appear to be at the early stage of galaxy evolution. These characteristics are analogous to some of the primeval galaxies which were formed in a primordial gas environment during the early stages of cosmic history, and provide an excellent laboratory in the local Universe for the studies of galaxy evolution and the formation of massive stars \citep{Bromm2009,Bromm2011,Wise2012,Fukushima2024}. 

Additionally, XMPs have been used to study Big Bang nucleosynthesis such as determining the primordial $^4$He abundance \citep{Fukugita2006,Izotov2014,Peimbert2016,Fernandez2019,Hsyu2020,Aver2021,Matsumoto2022}, because they have not experienced much chemical evolution. The relative abundances of the light elements (such as H, He, and Li) that were formed shortly after the Big Bang allow us to study the properties of the early Universe, as well as search for possible extensions to the Standard Model \citep{Steigman2007}. 

Given the great interest and wide applications of XMPs, there have been many attempts to search for new candidates. With great effort, the number of XMPs has been increased from 31 samples listed in the review of \citet{Kunth2000} to a few hundreds of XMPs reported in literature \citep[e.g. ][]{Kojima2020empress,Guseva2007,Guseva2017,Hirschauer2016,Hsyu2017,Hsyu2018,Izotov2006,Izotov2007,Izotov2009,IzotovThuan2009,Izotov2012,James2017,Pustilnik2010,Ruiz-Escobedo2018,Skillman2013,Thuan2005,vanZee2000,Yang2017}. The modern searches of XMP samples include the following approaches: (1) search for low redshift H\,\textsc{i} 21\,cm emission associated with blue optical colours \citep[e.g.][]{Skillman2013,Hirschauer2016,Karachentsev2023}; (2) human inspection and colour selection of multi-band imaging to identify galaxies that look similar to known XMP \citep[e.g][]{Hsyu2018,Grossi2025}; (3) trawling survey spectra to identify galaxies with weak metal emission lines \citep[e.g][]{Guseva2017,Zou2024}; and (4) applying machine learning algorithms to photometric properties to create a list of XMP candidates, combined with follow-up longslit spectroscopy \citep{Kojima2020empress}. Without spectroscopy, the identification process of (1) and (2) is generally based on colour-colour selections. The selection criteria usually also require human inspection. This not only leads to a prejudiced decision boundary, but one also cannot easily assess more than three dimensions of colour-colour diagrams with human inspection alone. 

Approach (4), by \citet{Kojima2020empress}, is the first attempt with a machine learning approach using a neural network classifier that can provide numerical decision boundaries in a multi-dimensional space. Their goal is to separate XMPs from stars, QSOs, and other galaxies using the photometric magnitudes in different bands such as the Hyper Suprime-Cam (HSC) $griz$ bands and the Sloan Digital Sky Survey (SDSS) $ugriz$ bands. The photometric magnitudes are generated from the spectral energy distribution (SED) models of star, QSOs, XMPs and non-XMPs for the HSC and SDSS, respectively. Hence, their training XMP samples are a set of emulated magnitudes from the SED models covering the physical properties of typical known XMPs. The samples for validation are selected at $z<0.03$ with \OH\ $<7.69$ ($\sim0.1Z_{\odot}$) from literatures. Their classifier accomplishes a completeness of 86\% and a purity of 46\% XMP classifications. This indicates that their classifier could possibly misidentify over a half of the samples as XMPs, given only the photometric magnitudes in different bands. 

In this work, we promote three main improvements with our deep learning (DL) pipeline composed of three individual models, made of convolutional neural networks (CNN). First, we simultaneously consider morphology and colour information given by multi-band imaging data, instead of using just photometric magnitudes. In addition to providing more information, the morphological and colour features are directly extracted by the CNNs without human interference. Our approach could be more inclusive towards different types of XMPs than applying a specific SED model. Secondly, due to the scarcity of the XMPs in the local Universe, it can be challenging to look for this tiny needle from a haystack of galaxies. Our DL pipeline performs a sequential classification procedure to simplify the task for each CNN model that helps to purify the XMP classification. Finally, in addition to classification, the pipeline predicts a proxy of metallicity that can be used to select the most promising XMP candidates. 

Our paper is outlined as follows: The datasets used in this work are described in Section~\ref{sec:dataset}, while the details of our methodology are introduced in Section~\ref{sec:method}. With the DL prediction, the selection of XMP candidates and the observations of these candidates are described in Section~\ref{sec:observation}. The analysis and discussion of the new observations are carried out in Section~\ref{sec:analysis}. We summarise the validation of the proposed methodology and the analyses of our new observations in Section~\ref{sec:summary}.

\section{Dataset}
\label{sec:dataset}
All DL approaches rely on reliable training dataset and labels in order to perform their task. In this section, we describe the dataset and labels that we have adopted to train our algorithm. For the labels, we used the N2 index (N2\,$\equiv\log$\{\rNii/\Ha\}) as the proxy for metallicity, probed by the oxygen abundance \citep{Storchi-Bergmann1994,Raimann2000,Denicolo2002,Pettini2004,Yin2007}, to select XMP candidates. The photometric data (including five bands, $u$, $g$, $r$, $i$, $z$) of the SDSS Data Release 17 \citep[DR17;][]{sdss17-2022} were used as the input images to the DL pipeline. To preserve colour information in the CNN models, we converted the fluxes (pixel values) of galaxy cutouts in each band into the relative fluxes to the ones of the $r$-band image by 

\begin{equation}
    \mathrm{f}_\mathrm{jk,norm} = \frac{\mathrm{f}_\mathrm{jk}-n_\mathrm{k}}{\mathrm{f}_{r,\mathrm{max}}-n_r},
\end{equation}
\noindent where the subscript j represents the pixel index from 0 to $N-1$ (with $N$ being the number of pixels), and the subscript k indicates 5 filters: $u$, $g$, $r$, $i$, and $z$. The numerator measures the intrinsic flux by subtracting the raw flux of each pixel in each band ($\mathrm{f}_\mathrm{jk}$) by the background level of each band ($n_\mathrm{k}$) which is determined by the average flux of the most common pixel values in a galaxy cutout of the filter. To determine the most common value, we construct a histogram of the pixel fluxes, and set the background level to be the peak of the distribution. 
The denominator represents the difference between the maximum value of flux in $r$-band ($\mathrm{f}_{r,\mathrm{max}}$) and the background level measured in $r$-band ($n_r$). The scaled flux of each pixel in each band, $\mathrm{f}_\mathrm{jk,norm}$, therefore provides an analogy of `colour' per pixel for each band relative to the $r$-band. 


\subsection{Known samples: training the CNN models}
\label{sec:training_samples}
With the SDSS spectroscopic observation, we select samples that not only have a spectroscopic measurement of the N2 index but also have appropriate imaging coverage across all five SDSS bands. The N2 index is provided by the MPA-JHU measurements -- \texttt{galSpec} -- using SDSS DR12 data\footnote{\url{https://www.sdss4.org/dr17/spectro/galaxy\_mpajhu/}} \citep{Kauffmann2003a,Brinchmann2004,Tremonti2004}. This gives the N2 index for 180\,369 galaxies. Additionally, we collect other known low-metallicity samples and their N2 values from the series of the EMPRESS project \citep{Kojima2020empress,Isobe2022empress,Nakajima2022empress,Xu2022ApJempress} and the following literature: \citet{Guseva2007,Guseva2017,Hirschauer2016,Hsyu2018,Izotov2006,Izotov2007,Izotov2009,IzotovThuan2009,James2017,Pustilnik2010,Ruiz-Escobedo2018,Skillman2013,Thuan2005,vanZee2000}. This provides the N2 index of an additional 108 galaxies. 

In this work, we define a metal-poor galaxy (hereafter MP) as a galaxy with $\mathrm{N2}\leq-1.0$, and an XMP is defined to have $\mathrm{N2}\leq-1.5$ [approximately \OH\ $\lessapprox8.0$; estimated using Equation~9 in \citet[][hereafter Y07]{Yin2007}]. With the collection of the SDSS spectroscopic samples and additional MP samples from the literature, the initial training sample contains 180\,477 galaxies in total with 5\,097 MP ($\mathrm{N2}\leq-1.0$; $\sim$2.82 per cent of total samples) and 384 XMP ($\mathrm{N2}\leq-1.5$; $\sim$0.2 per cent of total samples). 

\subsection{Training samples and labels}
\label{sec:label}

With the scarce number of XMPs, the prediction of the N2 index could be driven by the majority of samples with higher metallicities. Hence, we carry out a sequential approach with multiple CNN models focusing on different tasks: (i) classifying MP candidates from the total sample; (ii) classifying XMP candidates from the predicted MP samples; and (iii) predicting the N2 index from the predicted XMP samples (see details in Section~\ref{sec:training}). To perform these different tasks, the corresponding training samples and labels are different, as follows: 
\begin{enumerate}
    \item MP classifier: applied to all samples
    \begin{itemize}
        \item MP: $\mathrm{N2}\leq-1.0$
        \item non-MP: $\mathrm{N2}>-1.0$
    \end{itemize}
    \item XMP classifier: applied to the subset with $\mathrm{N2}\leq-0.5$
    \begin{itemize}
        \item XMP: $\mathrm{N2}\leq-1.5$
        \item non-XMP: $-0.5\geq\mathrm{N2}>-1.5$
    \end{itemize}
    \item N2 predictor: applied to the MP samples
\end{enumerate}
\noindent Amongst these samples, we randomly select a set of testing data, from the distribution of the training samples. These testing data are removed from the training procedures. To reduce the impact of data selection on training a CNN model, we also create three different training and testing sets for each procedure [(i), (ii), and (iii)]. 

Finally, to avoid any training bias caused by the number differences between the target outputs in each task \citep[examined in][]{Cheng2020a}, we balance the number of samples between the target outputs by rotating galaxy cutouts. Each galaxy is rotated by 90, 180, and 270 degrees due to the fixed cutout frame. This provides only three times more extra data in training. To further increase our training dataset, an additional negligible Gaussian noise, generated with a dispersion equal to 1 per cent of the standard deviation of the pixel values in a cutout, is then added to each cutout after rotation. With these tiny perturbations in inputs, the models are further improved even if using the repeated rotated images \citep[check the discussion of relevant approaches in][]{Goodfellow2014}. Note that in this work the utilisation of these additive noises is not to change the visual appearance of the cutout nor help regularisation of the model training, but just provide a nominal difference in pixel values. This step of data augmentation ensures that our training is performed on a balanced dataset. 

For the classifiers at the procedure (i) and (ii), the balancing is carried out between two target classes. For the N2 predictor at the procedure (iii), the target output is a floating value. Hence, we augment the data across several bins of the N2 index to ensure that each bin has an equal number of samples. We divide the range between $-1.0$ and $-2.1$ into 11 bins with an interval of 0.1. For the samples with $\mathrm{N2}\leq-2.1$, we form one bin due to the scarce population in this range. There are 2\,064 galaxies in the first bin of $\left [ -1.0,-1.1 \right )$, and the last bin of $\left [ -2.1, \right )$ contains 15 galaxies. The number of data in each bin is augmented to equal the number of data in the first bin.


\subsection{Working samples: SDSS images without spectroscopy}
\label{sec:target_samples}
\begin{table}
    \caption{The selection criteria for SDSS DR17 data query.}
    \begin{tabular}{cccc} 
        \hline
        \multicolumn{2}{c}{$16\leq{\mathrm{mag}_r}\leq22$} &
        \multicolumn{1}{c}{$(u-g)\leq1.7$} & {$(u-r)\leq2.1$}  \\ 
        \multicolumn{1}{c}{$(u-i)\leq2.2$} & {$(u-z)\leq2.5$} & {$(g-r)\leq0.6$} & {$(g-i)\leq0.9$} \\
        \multicolumn{1}{c}{$(g-z)\leq1.2$} & {$(r-i)\leq0.7$} & {$(r-z)\leq0.9$} & {$(i-z)\leq1.0$} \\
        \hline
    \end{tabular}
    \label{tab:selection_crit}
\end{table}

In this subsection, we describe the SDSS imaging sample that we use, in combination with our trained algorithm, to discover new XMP candidates that do not currently have spectroscopic confirmation from SDSS. The query for the SDSS DR17 is based on the physical properties such as colours and brightness of the known MP samples in the initial training samples. We applied query criteria as shown in Table~\ref{tab:selection_crit} when retrieving SDSS DR17 imaging data. The query criteria covers greater than 99\% of the MP samples in our dataset. Note that the colour criteria was applied to prevent wasting computational resources on unlikely samples, rather than determining the final list of candidates. We therefore allowed a broader coverage in these criteria, such that only the upper limits of the colour distributions were used, to avoid excluding desired samples that may have physical properties that differ from the currently known MPs.\footnote{We note that our final XMP candidates are all well-inside our selection box. This suggests our selection box has not significantly impacted the selection of XMP candidates.} Additionally, as CNN models are capable of extrapolating beyond the training distributions \citep{Cheng2021b,Cheng2023}, we considered galaxies that are up to 2 magnitudes fainter in $r$-band than the training set. The limit of $\mathrm{mag}_r\le22$ is chosen as it becomes challenging to observe any candidates with a continuum fainter than this limit using 4-meter telescopes (see Section~\ref{sec:observation}). By excluding data with SDSS spectroscopic observation and our samples, the number of working samples is 7\,763\,821. 


\section{Deep Learning Approach}
\label{sec:method}
\begin{table*}
    \begin{tabular}{rccccccccc} 
        \hline
        \multicolumn{1}{r}{} & {learning rate} & {l2} & {dropout} & {conv\_1} & {conv\_2} & {kernel\_1} & {kernel\_2} & {neuron\_1} & {neuron\_2}\\
        \hline\hline
        \multicolumn{1}{r}{MP classifier} & {0.0001} & {0.0} & {0.5} & {16} & {256} & {3} & {7} & {64} & {156}\\
        \multicolumn{1}{r}{XMP classifier} & {0.0001} & {0.0} & {0.0} & {256} & {256} & {3} & {3} & {512} & {16}\\
        \multicolumn{1}{r}{N2 predictor} & {0.0004} & {0.0} & {0.0} & {128} & {128} & {7} & {3} & {128} & {256}\\
        \hline
    \end{tabular}
    \caption{The hyper-parameters used for each CNN model. The `conv\_1' and `kernel\_1' are the channel and kernel size for the first convolutional layer (Conv 1), and the `conv\_2' and `kernel\_2' are for the second convolutional layer (Conv 2). The `neuron\_1' and `neuron\_2' are the number of neurons used in the dense layers (Dense 1 and Dense 2), respectively.}
    \label{tab:archi_hyper}
\end{table*}
\begin{figure}{}
\begin{center}
\graphicspath{}
	\includegraphics[width=\columnwidth]{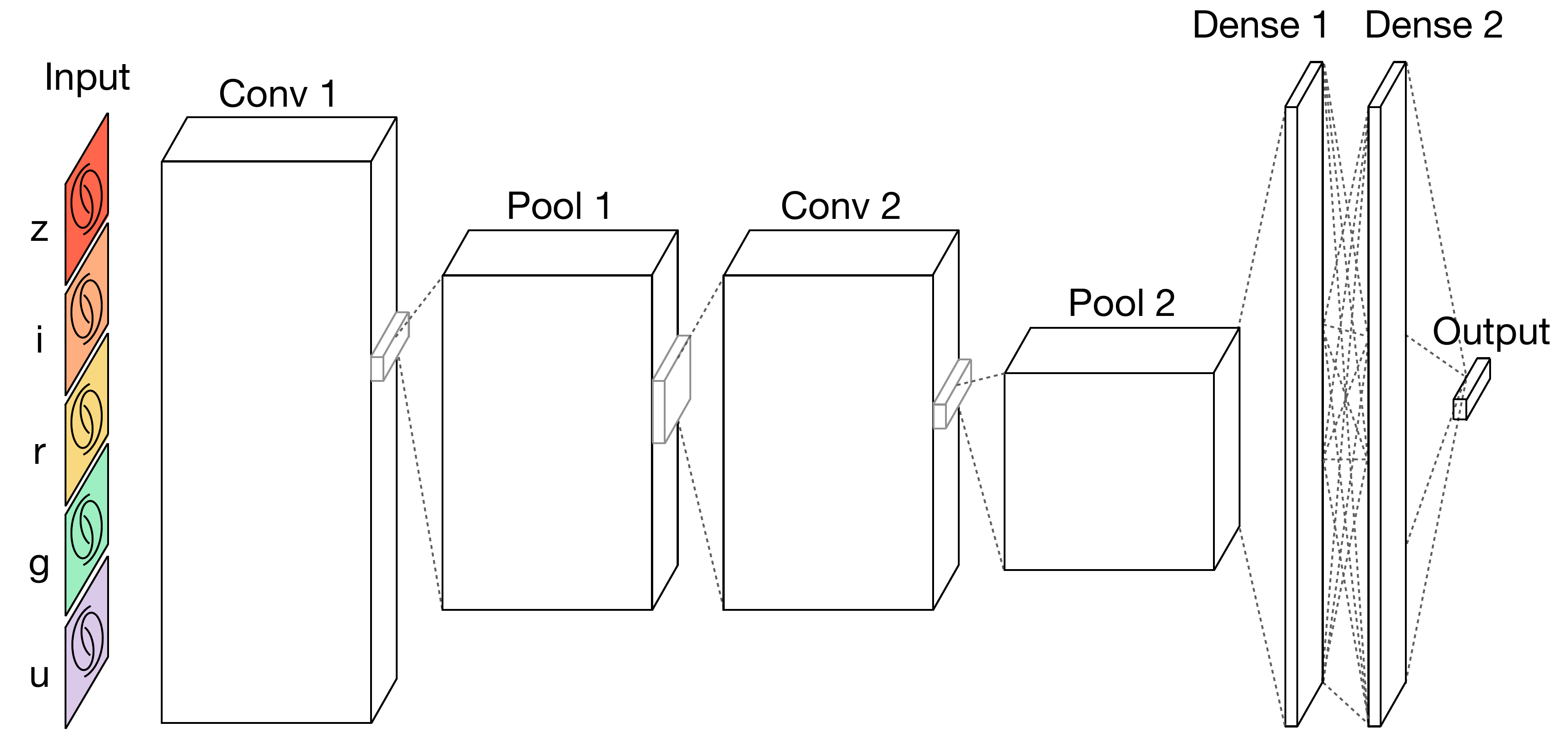}
   	\caption{Schematic diagram of the CNN architecture used in this work. The input is a galaxy image of 5 different filters ($u$, $g$, $r$, $i$, $z$). The `Conv 1' and `Conv 2' represent convolutional layers, and each layer is followed by a pooling layer (Pool 1 and Pool 2), respectively. Finally, two dense layers (Dense 1 and Dense 2) are used before the output layer.}
    \label{fig:cnn_architecture}
\end{center}
\end{figure}
We use multi-band imaging data as the input of a DL pipeline to carry out sequential predictions of classifying XMP candidates and estimating their N2 index. Due to the scarce population of XMPs, the DL pipeline is composed of three CNN algorithms: (i) MP classifier; (ii) XMP classifier; and (iii) N2 predictor. The architecture of each CNN model is the same (see Fig.~\ref{fig:cnn_architecture}) for simplicity. The input cutouts have a dimension of 32 by 32 pixels and contain 5 bands ($u$, $g$, $r$, $i$, $z$). The architecture contains two convolutional layers (Conv 1 and Conv 2) followed by a pooling layer (Pool 1 and Pool 2) for each convolutional layer, as well as two dense layers (Dense 1 and Dense 2). Two dropout layers were implemented after the second pooling layer and before the output layer, to reduce the number of parameters inside the networks. The dropout rate is one of the hyper-parameters and a constant for both dropout layers. 

The hyper-parameters for each CNN model, however, are optimised individually for each model using the Bayesian optimisation method \citep{Frazier2018} due to the use of different datasets (see Section~\ref{sec:label}) in training different models. Table~\ref{tab:archi_hyper} shows the hyper-parameters used for the MP classifier, the XMP classifier, and the N2 predictor, respectively. We applied the {\texttt{Adam}} optimiser \citep{Kingma2015}, and the learning rates are also hyper-parameters optimised independently for each model. The maximum number of iterations for each training is 20 epochs, but only the model with the minimum validation loss within the 20 epochs is saved.

\subsection{Training XMP classification and N2 prediction}
\label{sec:training}
\begin{figure*}{}
\begin{center}
\graphicspath{}
	\includegraphics[width=2.1\columnwidth]{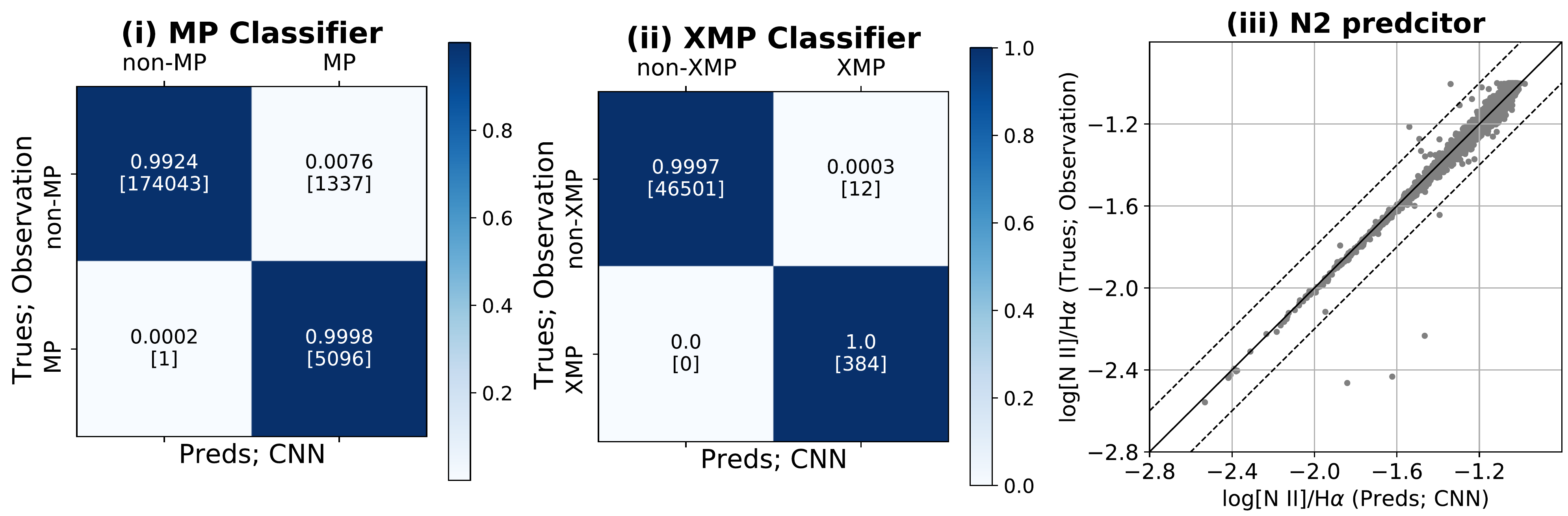}
   	\caption{Each panel presents the result applying the 9 trained CNN models to the specific dataset (including training and testing samples) for each procedure (Section~\ref{sec:label}). The left and middle panels are the confusion matrices of the MP and XMP classifiers. The classification probability thresholds for assigning classes are $>0.5$. The value in each quadrant indicates the fraction (number) of the samples predicted by CNN in each true (observed) class. The right panel shows the comparison between the predicted N2 index by CNN models and the observed N2 values from literature (listed in Section~\ref{sec:training_samples}). The solid line shows a one-to-one relation and the dashed lines indicate a scatter of 0.2 dex.
    }
    \label{fig:individual_result}
\end{center}
\end{figure*}

As mentioned in Section~\ref{sec:label}, we created three different training and testing sets for each procedure, (i), (ii), (iii) to account for the impact of the quality of randomly selected datasets. Furthermore, we train three independent CNN models for each procedure to account for the variation caused by having a random initial state when training a new model. Therefore, there are $3\times3$, i.e. 9 CNN models trained for each procedure, and each model is assessed by its corresponding testing set.

In this section, we describe the details of how each algorithm is trained and evaluated independently. For classifiers, the median values of the output probabilities from the 9 CNN models, trained for each classifier, are used to assign classes. Similarly, for the N2 predictor, the median value of the predicted N2 indices from the 9 models is used for the selection of the candidates. The validation of each procedure is carried out separately using their whole assigned samples (including training and testing sets) for each procedure, as stated in Section~\ref{sec:label}. 

In detail, the MP classifier is trained with all available samples (excluding their testing sets) separated into two classes: MP ($\mathrm{N2}\leq-1.0$) and non-MP ($\mathrm{N2}>-1.0$). The numbers of MP and non-MP samples in the training set are equal after data augmentation for training. The median value of the predicted probabilities from the 9 models is used to assign classes. The trained models are applied to all samples in this work including 180\,477 galaxies, and the result is shown at the left panel of the Fig.~\ref{fig:individual_result}. This confusion matrix of the MP classifier uses a threshold of $0.5$ applied to the median predicted probabilities. The classification accuracy is about 99.26\%, which indicates the fraction between the number of correctly classified samples and the total number of samples. The trained MP classifier correctly identifies over 99.98\% MP galaxies (i.e. the recall is 99.98\%). The false positive, which the model identifies as a MP but with $\mathrm{N2}>-1.0$, occupies about 20.78\% of the predicted MP samples, containing mostly (over a fraction of 0.9686) galaxies with $\mathrm{N2}\leq-0.5$. 

With the high fraction of false positives with $\mathrm{N2}\leq-0.5$, the XMP classifier is trained on the subset of samples with $\mathrm{N2}\leq-0.5$. The negative label, `non-XMP', is therefore for the galaxies with N2 between $-0.5$ and $-1.5$ (see also, Section~\ref{sec:label}). The numbers of XMP and non-XMP training samples are balanced with data augmentation for training. Again, the median value of the predicted probabilities is used to assign classes, and the middle panel of Fig.~\ref{fig:individual_result} shows the result applying the trained models of the XMP classifier to all subset samples with $\mathrm{N2}\leq-0.5$. The XMP classifier also reaches a high classification accuracy of 99.97\%. There are around 3\% of false positives, which the model identifies as a XMP but with $\mathrm{N2}>-1.5$. In this test, all of the false positives are in fact MP galaxies with $\mathrm{N2}\leq-1.0$. 

We therefore anticipate that the vast majority of the classified XMPs after the sequential classification of two classifiers shall satisfy the definition of MP. The N2 predictor is trained with only the MP samples with $\mathrm{N2}\leq-1.0$ to focus on effectively predicting the N2 index at the lowest range. As stated in Section~\ref{sec:label}, we separate the MP samples into 12 bins based on their N2 index. The first 11 bins have an interval of 0.1, and the last bin covers the remaining samples with a broader range of N2 values, $\left (-2.6 , -2.1 \right ]$, where $-2.6$ is the lowest N2 value in our sample. We augment the number of data in each bin to match the number of data in the first bin, $\left (-1.1, -1.0 \right ]$. Since the primary goal of this work is to identify XMP candidates with the lowest possible N2 index, we introduce a loss weighting factor ($3\times$) for the systems that have a true N2 index $\le-2.1$. This ensures that the network is more severely penalised when it incorrectly predicts the N2 index of the most metal-poor XMPs.

The right panel of Fig.~\ref{fig:individual_result} shows the comparison of all MP samples (5\,097 galaxies) between the predicted N2 index by CNN and the observed values collected from the literature (Section~\ref{sec:training_samples}). The prediction of the N2 index is accurate with the root-mean-squared deviation (RMSD) of 0.031 dex and the median absolute deviation (MAD) of 0.015 dex.

\subsection{Evaluation of the sequential process}
\label{sec:assessment_sequential}
\begin{figure*}{}
\begin{center}
\graphicspath{}
	\includegraphics[width=1.5\columnwidth]{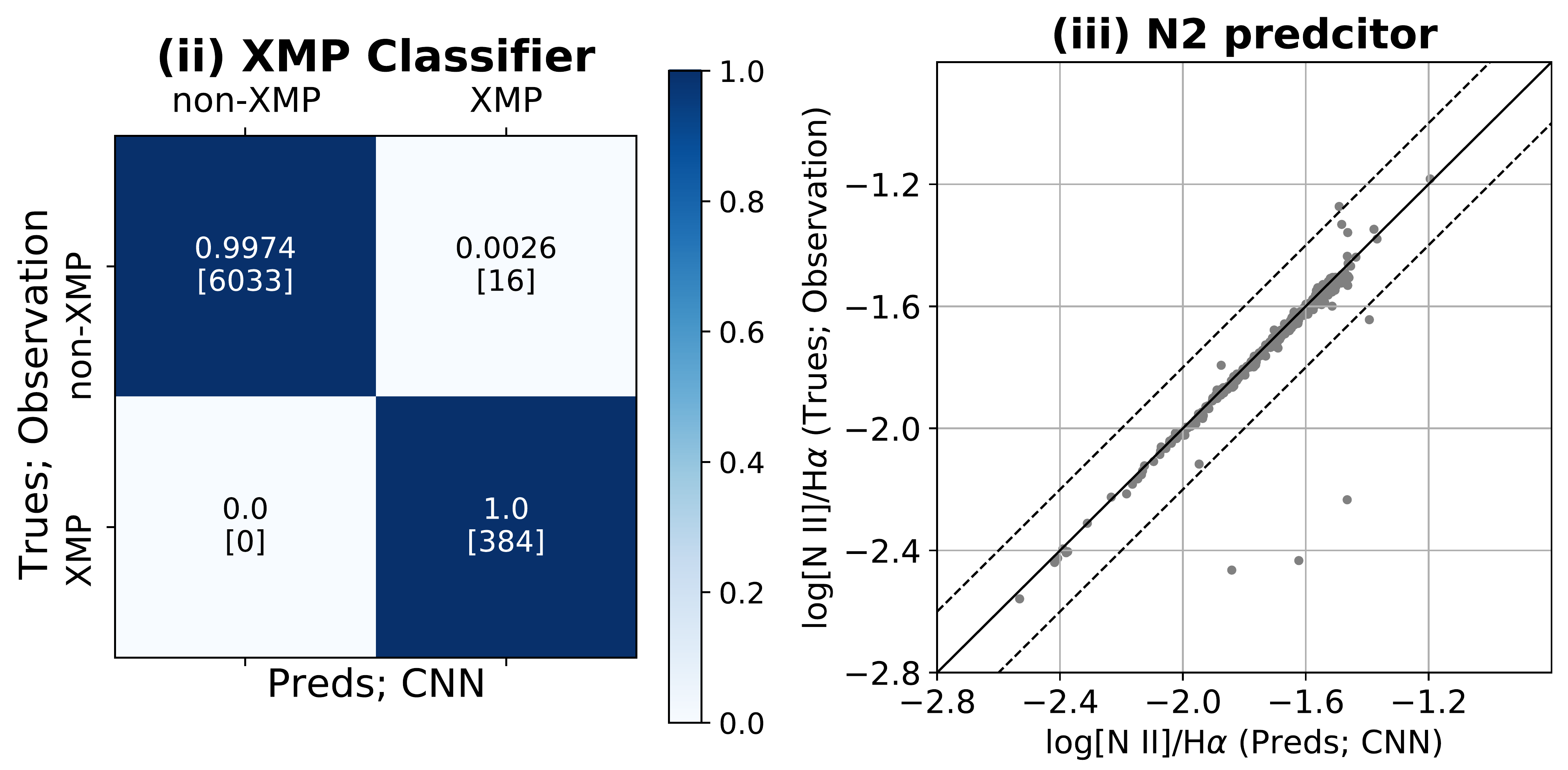}
   	\caption{Unlike Fig.~\ref{fig:individual_result}, this figure presents the evaluation of the sequential process. The left panel shows the confusion matrix of the XMP classifier for the predicted MP candidates from Fig.~\ref{fig:individual_result}. The classification probability threshold is $>0.5$. The value in each quadrant indicates the fraction (number) of the samples predicted by CNN in each true (observed) class. The right panel shows the comparison between the predicted N2 index of the predicted XMP candidates by CNN models and their observed N2 values from literature. The solid line shows a one-to-one relation and the dashed lines indicate a scatter of 0.2 dex.
    }
    \label{fig:sequential_result}
\end{center}
\end{figure*}
When a new set of galaxy cutouts is fed to the sequential process, only the predicted MP candidates with $P_{\mathrm{MP}}>0.5$ from the MP classifier proceed to the XMP classifier; similarly, only the predicted XMP candidates with $P_{\mathrm{XMP}}>0.5$ from the XMP classifier advance to the N2 predictor. Therefore, the sequential process constructs a list of XMP candidates with their MP predicted probability ($P_{\mathrm{MP}}>0.5$), XMP predicted probability ($P_{\mathrm{XMP}}>0.5$), and the predicted N2 index. The output N2 index is used to select the most promising candidate for observation, which we discuss further in Section~\ref{sec:selection}.

The assessment of the sequential process with all available samples (i.e. 180\,477 galaxies) is shown in Fig.~\ref{fig:sequential_result}. Unlike Fig.~\ref{fig:individual_result}, only the predicted MP candidates in the left panel of Fig.~\ref{fig:individual_result} (i.e. 6\,433 samples) proceed to the XMP classifier (the left panel of Fig.~\ref{fig:sequential_result}), and only the predicted XMP candidates (containing 400 samples) continue to the N2 predictor (the right panel of Fig.~\ref{fig:sequential_result}). With the two classifiers carrying out a sequential classification, the fraction of 0.96 and 0.99 of predicted XMP candidates are indeed XMP galaxies and MP galaxies, respectively (i.e. the precision is 96\% and 99\%). The predicted N2 index has the RMSD of 0.13 dex and a MAD of 0.012 dex. With such small statistics, the RMSD is skewed towards the outliers; while the MAD, which is less affected by outliers, is consistent with the individual test in Fig.~\ref{fig:individual_result}.

\section{Observations and Data Reduction}
\label{sec:observation}
\begin{table*}
    \begin{center}
    \caption{The list of observed XMP samples from INT and SOAR telescope. The order is sorted based on the predicted N2 values by our CNN algorithms. The `ObsDate' is the date that the observations were conducted. The `ExpTime' provides the exposure time used for a single exposure, and the `ExpN' is the number of exposures for an object. The `ref.' column provides references to XMP galaxies that have been investigated in the literature.}
    \label{tab:obs_samples}
    \begin{tabular}{lcccccccc} 
        \hline
        \multicolumn{1}{c}{Name$^{a}$} & {RA} & {DEC} & {N2} & {Instrument} & {ObsDate} & {ExpTime} & {ExpN} & {ref.$^{b}$}\\
         & {(J2000.0)} & {(J2000.0)} & {(CNN)} & {} & {} & {(s)} & {} & {}\\        
        \hline\hline
        \multicolumn{1}{l}{XMP0013+1354} & {00 13 42.62} & {+13 54 17.1} & {-1.92$\pm$0.07} & {INT/IDS} & {21-Aug-2023} & {1000} & {3} & {$\ldots$}\\
        \multicolumn{1}{l}{XMP0124+0838} & {01 24 06.58} & {+08 38 06.9} & {-2.17$\pm$0.10} & {SOAR/Goodman} & {8-Nov-2023} & {800} & {3} & {$\ldots$}\\
        \multicolumn{1}{l}{XMP0219$-$0059} & {02 19 30.34} & {$-$00 59 14.3} & {-1.81$\pm$0.04} & {SOAR/Goodman} & {6-Oct-2023} & {800} & {3} & {1,7}\\
        \multicolumn{1}{l}{XMP0429+0026} & {04 29 51.38} & {+00 26 52.0} & {-1.88$\pm$0.03} & {INT/IDS} & {4-Mar-2024} & {1200} & {3} & {$\ldots$}\\
        \multicolumn{1}{l}{XMP0742+1103} & {07 42 18.07} & {+11 03 30.6} & {-1.86$\pm$0.04} & {INT/IDS} & {5-Mar-2024} & {1800} & {3} & {$\ldots$}\\
        \multicolumn{1}{l}{XMP0752+2340} & {07 52 11.06} & {+23 40 16.7} & {-2.00$\pm$0.09} & {INT/IDS} & {3-Mar-2024} & {1000} & {3} & {$\ldots$}\\
        \multicolumn{1}{l}{XMP0801+2640} & {08 01 03.92} & {+26 40 54.3} & {-1.89$\pm$0.05} & {INT/IDS} & {5-Mar-2024} & {600} & {3} & {2}\\
        \multicolumn{1}{l}{XMP0803+1635} & {08 03 16.04} & {+16 35 44.6} & {-1.81$\pm$0.06} & {INT/IDS} & {3-Mar-2024} & {1500} & {3} & {$\ldots$}\\
        \multicolumn{1}{l}{XMP0827+1059} & {08 27 46.65} & {+10 59 11.1} & {-1.97$\pm$0.06} & {INT/IDS} & {6-Mar-2024} & {1500} & {3} & {12}\\
        \multicolumn{1}{l}{XMP0850+1150} & {08 50 57.57} & {+11 50 45.6} & {-2.06$\pm$0.07} & {INT/IDS} & {28-Feb-2024} & {1000} & {3} & {$\ldots$}\\
        \multicolumn{1}{l}{XMP0856+2414} & {08 56 01.07} & {+24 14 21.5} & {-1.88$\pm$0.04} & {INT/IDS} & {6-Mar-2024} & {1400} & {3} & {6}\\
        \multicolumn{1}{l}{XMP0916+5002} & {09 16 06.66} & {+50 02 30.6} & {-2.11$\pm$0.07} & {INT/IDS} & {5-Mar-2024} & {1800} & {3} & {$\ldots$}\\
        \multicolumn{1}{l}{XMP0916+0257} & {09 16 25.09} & {+02 57 43.2} & {-2.13$\pm$0.02} & {INT/IDS} & {4-Mar-2024} & {1800} & {3} & {$\ldots$}\\
        \multicolumn{1}{l}{XMP0922+6324} & {09 22 23.86} & {+63 24 36.9} & {-2.12$\pm$0.06} & {INT/IDS} & {4-Mar-2024} & {2000} & {3} & {12}\\
        \multicolumn{1}{l}{XMP0928+3601} & {09 28 44.73} & {+36 01 04.2} & {-1.91$\pm$0.08} & {INT/IDS} & {3-Mar-2024} & {1200} & {3} & {4}\\
        \multicolumn{1}{l}{XMP0930+4934} & {09 30 04.97} & {+49 34 29.7} & {-1.95$\pm$0.06} & {INT/IDS} & {28-Feb-2024} & {1000} & {3} & {$\ldots$}\\
        \multicolumn{1}{l}{XMP0931+2617} & {09 31 14.14} & {+26 17 27.4} & {-1.98$\pm$0.15} & {INT/IDS} & {6-Mar-2024} & {1800} & {3} & {$\ldots$}\\
        \multicolumn{1}{l}{XMP1003+2746} & {10 03 10.80} & {+27 46 31.7} & {-1.92$\pm$0.13} & {INT/IDS} & {6-Mar-2024} & {1600} & {3} & {$\ldots$}\\
        \multicolumn{1}{l}{XMP1030+3151} & {10 30 44.81} & {+31 51 24.0} & {-2.04$\pm$0.09} & {INT/IDS} & {3-Mar-2024} & {1100} & {3} & {$\ldots$}\\
        \multicolumn{1}{l}{XMP1032+5035} & {10 32 00.39} & {+50 35 07.7} & {-2.08$\pm$0.08} & {INT/IDS} & {5-Mar-2024} & {1000} & {3} & {$\ldots$}\\
        \multicolumn{1}{l}{XMP1035+3814} & {10 35 07.20} & {+38 14 30.4} & {-1.89$\pm$0.02} & {INT/IDS} & {28-Feb-2024} & {1500} & {3} & {6}\\
        \multicolumn{1}{l}{XMP1139+0040} & {11 39 00.41} & {+00 40 42.6} & {-2.02$\pm$0.07} & {INT/IDS} & {4-Mar-2024} & {1500} & {3} & {12}\\
        \multicolumn{1}{l}{XMP1140+5037} & {11 40 45.72} & {+50 37 07.6} & {-1.95$\pm$0.06} & {INT/IDS} & {5-Mar-2024} & {1400} & {3} & {$\ldots$}\\
        \multicolumn{1}{l}{XMP1214+1245} & {12 14 33.11} & {+12 45 49.2} & {-2.18$\pm$0.02} & {INT/IDS} & {3-Mar-2024} & {1100} & {3} & {3}\\
        \multicolumn{1}{l}{XMP1228+4313$^*$} & {12 28 48.09} & {+43 13 48.9} & {-2.00$\pm$0.04} & {INT/IDS} & {6-Mar-2024} & {600} & {3} & {$\ldots$}\\
        \multicolumn{1}{l}{XMP1230+0544} & {12 30 11.99} & {+05 44 50.7} & {-1.96$\pm$0.02} & {INT/IDS} & {6-Mar-2024} & {1000} & {3} & {$\ldots$}\\
        \multicolumn{1}{l}{XMP1238+3246$^*$} & {12 38 40.25} & {+32 46 00.9} & {-2.02$\pm$0.02} & {INT/IDS} & {28-Feb-2024} & {2200} & {3} & {10}\\
        \multicolumn{1}{l}{XMP1322+2251} & {13 22 01.75} & {+22 51 31.5} & {-2.03$\pm$0.03} & {INT/IDS} & {5-Mar-2024} & {1800} & {2} & {$\ldots$}\\
        \multicolumn{1}{l}{XMP1329+2237} & {13 29 24.31} & {+22 37 12.3} & {-2.00$\pm$0.05} & {INT/IDS} & {28-Feb-2024} & {1000} & {3} & {$\ldots$}\\
        \multicolumn{1}{l}{XMP1344+0621} & {13 44 57.48} & {+06 21 46.3} & {-1.96$\pm$0.02} & {INT/IDS} & {6-Mar-2024} & {1000} & {3} & {$\ldots$}\\
        \multicolumn{1}{l}{XMP1347+0755} & {13 47 56.00} & {+07 55 32.1} & {-2.31$\pm$0.07} & {INT/IDS} & {4-Mar-2024} & {1500} & {3} & {12}\\
        \multicolumn{1}{l}{XMP1408+1753} & {14 08 16.16} & {+17 53 50.9} & {-1.86$\pm$0.04} & {INT/IDS} & {3-Mar-2024} & {1000} & {3} & {$\ldots$}\\
        \multicolumn{1}{l}{XMP1422+5414} & {14 22 38.85} & {+54 14 09.2} & {-1.81$\pm$0.07} & {INT/IDS} & {3-Mar-2024} & {600} & {3} & {8}\\
        \multicolumn{1}{l}{XMP1631+4426} & {16 31 14.25} & {+44 26 04.7} & {-2.42$\pm$0.02} & {INT/IDS} & {4-Mar-2024} & {1800} & {3} & {5,9}\\
        \multicolumn{1}{l}{XMP1638+2421} & {16 38 58.01} & {+24 21 39.2} & {-2.07$\pm$0.08} & {INT/IDS} & {5-Mar-2024} & {1200} & {3} & {$\ldots$}\\
        \multicolumn{1}{l}{XMP1655+6337} & {16 55 39.20} & {+63 37 03.3} & {-2.10$\pm$0.02} & {INT/IDS} & {6-Mar-2024} & {1500} & {1} & {3}\\
        \multicolumn{1}{l}{XMP2048$-$0559} & {20 48 34.22} & {$-$05 59 01.4} & {-2.07$\pm$0.08} & {INT/IDS} & {21-Aug-2023} & {2300} & {1} & {$\ldots$}\\
        \multicolumn{1}{l}{XMP2136$-$0307} & {21 36 09.38} & {$-$03 07 30.7} & {-1.86$\pm$0.05} & {SOAR/Goodman} & {6-Oct-2023} & {800} & {3} & {$\ldots$}\\
        \multicolumn{1}{l}{XMP2149$-$0535} & {21 49 12.62} & {$-$05 35 05.6} & {-2.16$\pm$0.08} & {INT/IDS} & {21-Aug-2023} & {3000} & {1} & {$\ldots$}\\
        \multicolumn{1}{l}{XMP2156+0856} & {21 56 33.58} & {+08 56 36.6} & {-1.86$\pm$0.06} & {INT/IDS} & {20-Aug-2023} & {800} & {2} & {6}\\
        \multicolumn{1}{l}{XMP2212+2205} & {22 12 59.31} & {+22 05 05.5} & {-1.89$\pm$0.03} & {INT/IDS} & {21-Aug-2023} & {1000} & {3} & {$\ldots$}\\
        \multicolumn{1}{l}{XMP2325+2008} & {23 25 37.75} & {+20 08 17.2} & {-1.91$\pm$0.14} & {INT/IDS} & {22-Aug-2023} & {1000} & {3} & {$\ldots$}\\
        \multicolumn{1}{l}{XMP2329+0226} & {23 29 26.60} & {+02 26 28.1} & {-1.85$\pm$0.05} & {INT/IDS} & {21-Aug-2023} & {1000} & {3} & {11}\\
        \multicolumn{1}{l}{XMP2331+2226} & {23 31 00.91} & {+22 26 34.8} & {-1.81$\pm$0.04} & {INT/IDS} & {21-Aug-2023} & {2000} & {3} & {$\ldots$}\\
        \multicolumn{1}{l}{XMP2336$-$0404} & {23 36 31.64} & {$-$04 04 36.6} & {-1.96$\pm$0.08} & {SOAR/Goodman} & {8-Nov-2023} & {800} & {3} & {$\ldots$}\\
        \hline
    \end{tabular}
    \end{center}
    {\raggedright
    \scriptsize 
    $^a$ A `$*$' indicates a nearby H{\textsc{ii}} region ($z<0.002$). \\
    $^b$ Reference: (1) \citet{Ann2015}, (2) \citet{Griffith2011}, (3) \citet{Hsyu2018}, (4) \citet{James2017}, (5) \citet{Kojima2020empress}, (6) \citet{Liu2023}, (7) \citet{Micheva2013}, (8) \citet{Thuan1995}, (9) \citet{Thuan2022}, (10) \citet{vanZee2006} (H\textsc{ii} region), (11) \citet{Wang2018}, (12) \citet{Yang2017}. \\}
\end{table*}

\subsection{Sample Selection}
\label{sec:selection}
The trained CNN models are applied to the working samples (Section \ref{sec:target_samples}) -- SDSS DR17 multi-band imaging without SDSS spectroscopy. This gives 232\,954 XMP candidates with $P_\mathrm{MP}>0.5$ and $P_\mathrm{XMP}>0.5$ and the predictions of their N2 values from over 7 million SDSS galaxy cutouts. 

We select only the most promising XMP candidates with the lowest range of metallicities for observation by applying $P_\mathrm{MP}>0.99$, $P_\mathrm{XMP}>0.99$, and $\mathrm{N2}<-1.8$. We performed a fast visual inspection to exclude apparent artefacts and faulty images, leaving 390 highly possible XMP candidates with lowest ranges of N2 index ($<-1.8$) predicted by the DL pipeline. 
A few subtle artefacts remain among the samples, requiring further assessment before they can be selected as final observational candidates.

\subsection{Observations}
To validate the effectiveness of the DL pipeline as well as discover new XMP galaxies, we conducted observational programmes using the 2.54~m Isaac Newton Telescope (INT) and the 4.1~m Southern Astrophysical Research (SOAR) Telescope between 2023 and 2024. Table~\ref{tab:obs_samples} lists 45 targets for which we acquired spectra as part of these observations. The order is sorted according to their right ascension. If a galaxy has previously been observed in the literature, we also list its corresponding references. 
The details of each of these observing programmes are provided in the following subsections.

\subsubsection{INT observations}
\label{sec:int}
We collected 41 new optical spectra using the Intermediate Dispersion Spectrograph (IDS) with the \texttt{Red+2} detector and \texttt{R632V} grating on the INT during 19 August -- 22 August 2023 (2023B) and 28 February -- 6 March 2024 (2024A). The \texttt{Red+2} detector provides a spatial scale of $0.44$ arcsec per pixel. The detector binning was set to be $1\times1$ (i.e. unbinned). The \texttt{R632V} grating has a total wavelength coverage of 2178\AA\ and the spectral resolution of $R\sim2270$ at 4500\AA, with a nominal sampling of $\sim1$\AA/pixel. This grating was chosen to resolve the weak \Nii\ line from the much stronger \Ha\ line, as well as its better efficiency ($>60$\%) within the target range of wavelengths. All observations were made using a $1^{\prime\prime}$ slit, and the slit angle is at the approximate parallactic angle during the observations. We use observations of G191-B2B\footnote{\url{https://www.eso.org/sci/observing/tools/standards/spectra/g191b2b.html}} to flux calibrate the data taken in both observation periods. 

The 2023B observation serves as a preliminary assessment for validating the DL pipeline. Since the redshifts of our selected samples were unknown, a central wavelength of $6460.5$\AA\ was chosen to cover a wavelength range from 5370\AA\ to 7540\AA\ for good-quality observation of the \rNii\ line at $z<0.15$. Spectroscopic observation of 8 XMP candidates were made, and their redshifts were all less than 0.055. This observation provides an important validation of the DL pipeline. For the 2024A observation, we adjust the central wavelength to $5940$\AA\ with the prior knowledge from previous observation to ensure coverage on \Oiii\ doublet at $\lambda\lambda4960,5008\,$\AA, \Hb, \Nii\ doublet at $\lambda\lambda6550,6585\,$\AA, and \Ha\ lines. We observed 33 XMP candidates total, and their redshifts are all less than 0.065.

\subsubsection{SOAR observations}
\label{sec:soar}
We collect 4 spectra using the Goodman High Throughput Spectrograph \citep[Goodman;][]{Clemens2004} with the \texttt{SOAR\_GHTS\_BLUECAM} camera on the SOAR telescope on 6 October 2023 and 8 November 2023. The spatial scale is 0.15 arcsec per pixel, and the binning for the detector is set to be 2$\times$2. We use the \texttt{SYZY\_400} grating, which has a spectral resolution of $R\sim850$ at 5500\AA\ (assuming a 1 arcsec slit). This setup provides optimal throughout for the wavelength range covering the \Oiii\ doublet at $\lambda\lambda4960,5008\,$\AA, \Hb, \Nii\ doublet at $\lambda\lambda6550,6585\,$\AA, and \Ha\ lines. The slit size is set to 1$^{\prime\prime}$, and the slit angle is at the approximate parallactic angle during the observations. The standard star for flux calibration is LTT 3864\footnote{\url{https://www.eso.org/sci/observing/tools/standards/spectra/ltt3864.html}}. Three observations of 800 seconds each are made for each object.

\subsection{Data Reduction}
\label{sec:result}
\begin{figure*}{}
\begin{center}
\graphicspath{}
	\includegraphics[width=\columnwidth]{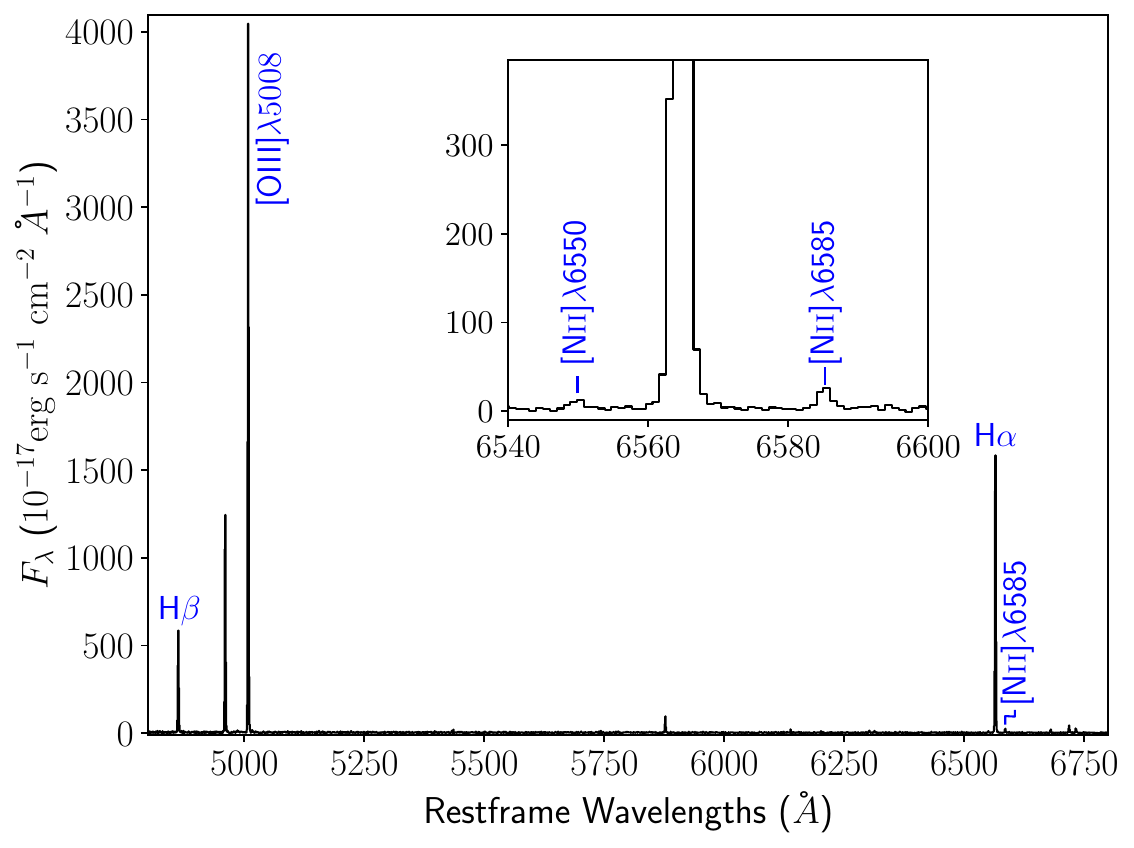}
        \includegraphics[width=\columnwidth]{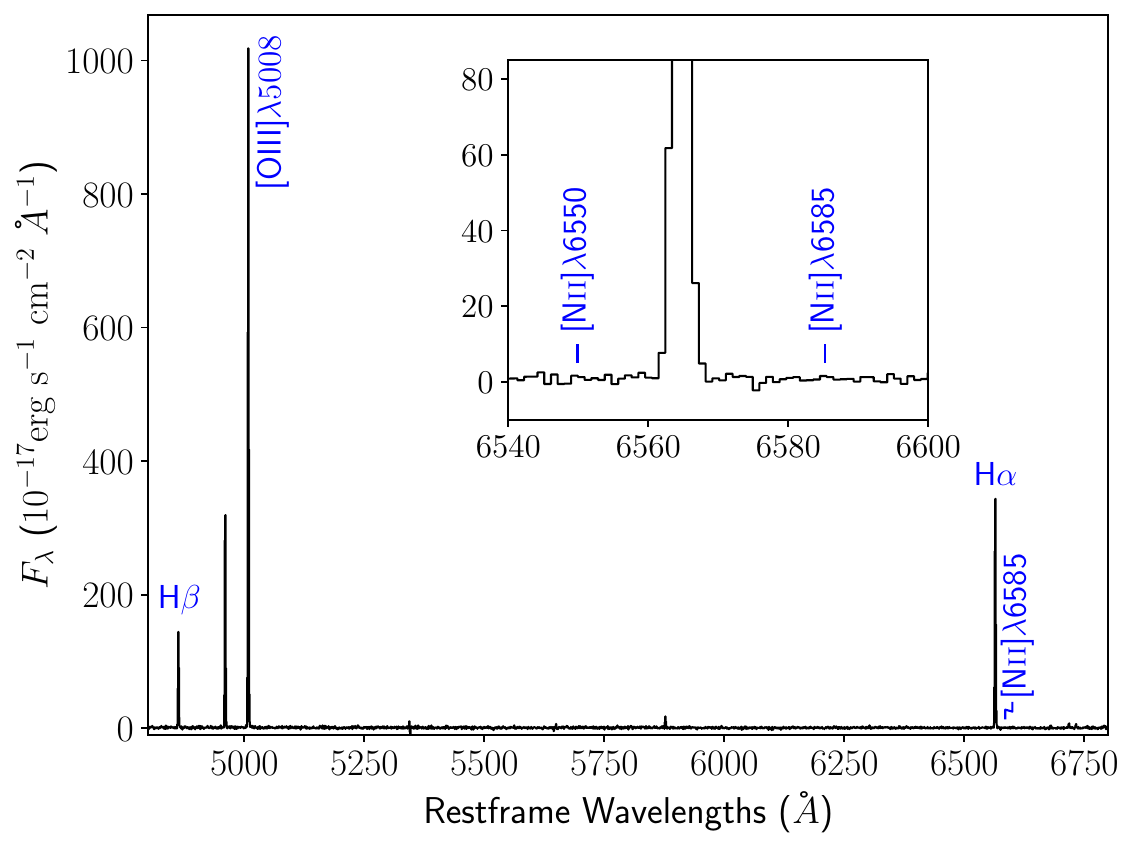}
   	\caption{Examples of the reduced spectra with (left; $>10\sigma$) and without (right; $<2\sigma$) clear detection of the \rNii\ lines.}
    \label{fig:example_spec}
\end{center}
\end{figure*}
The data reduction is carried out using the \textsc{PypeIt} data reduction pipeline \citep{pypeit:zenodo,pypeit:joss_pub}. The reduction process includes the subtraction of bias frames, the correction of the flat-field using dome flats, identification and masking of cosmic rays, sky subtraction, wavelength calibration using arc frames, 1D boxcar extraction and flux calibration using the chosen standard stars. When available, three exposures of a single candidate are combined using the \textsc{PypeIt} coaddition tools. The number of exposures for each target is also listed in Table~\ref{tab:obs_samples}. Examples of the reduced and combined spectra with clear (left) and unclear (right) detection of \rNii\ lines are shown in Fig.~\ref{fig:example_spec}. 

\section{Analysis and Discussion}
\label{sec:analysis}
\begin{table*}
    \hspace*{\fill}
    \begin{center}
    \caption{The measurements of H$\alpha$ emission line flux, the N2 index, and the O3 index. The missing values are indicated with three dots. For samples with significance $\geq2$, we provide the N2 values measured with the integrated flux approach; for other samples, we provide the $2\sigma$ values.}
    \label{tab:lines}
    \begin{tabular}{lccccc}
        \hline
        \multicolumn{1}{c}{Name} & {redshift} & {$F\left (\mathrm{H}\alpha \right)$$^{a}$} & {Significance$^{b}$} & {N2} & {O3} \\
        \hline\hline
        \multicolumn{1}{l}{XMP0013+1354} & {0.0524} & {2921$\pm$48} & {2.41} & {-1.99$\pm$0.21} & {$\ldots$}\\
        \multicolumn{1}{l}{XMP0124+0838} & {0.0487} & {912.7$\pm$4.4} & {9.47} & {-2.01$\pm$0.05} & {0.828$\pm$0.005}\\
        \multicolumn{1}{l}{XMP0219$-$0059} & {0.0085} & {1947.6$\pm$6.6} & {17.28} & {-2.24$\pm$0.04} & {0.716$\pm$0.004}\\
        \multicolumn{1}{l}{XMP0429+0026} & {0.0119} & {519.1$\pm$8.0} & {3.88} & {-2.04$\pm$0.18} & {0.801$\pm$0.015}\\
        \multicolumn{1}{l}{XMP0742+1103} & {0.0438} & {136.5$\pm$4.3} & {1.84} & {$<$-1.50} & {0.726$\pm$0.026}\\
        \multicolumn{1}{l}{XMP0752+2340} & {0.0474} & {1048$\pm$11} & {5.37} & {-1.93$\pm$0.09} & {0.910$\pm$0.007}\\
        \multicolumn{1}{l}{XMP0801+2640} & {0.0265} & {3963$\pm$27} & {12.98} & {-1.82$\pm$0.04} & {0.805$\pm$0.007}\\
        \multicolumn{1}{l}{XMP0803+1635} & {0.0211} & {440.4$\pm$6.0} & {2.71} & {-2.13$\pm$0.19} & {0.842$\pm$0.013}\\
        \multicolumn{1}{l}{XMP0827+1059} & {0.0436} & {327.4$\pm$7.6} & {2.27} & {-2.48$\pm$0.89} & {0.723$\pm$0.022}\\
        \multicolumn{1}{l}{XMP0850+1150} & {0.0293} & {620.9$\pm$8.9} & {2.37} & {-1.68$\pm$0.09} & {0.820$\pm$0.014}\\
        \multicolumn{1}{l}{XMP0856+2414} & {0.0511} & {295.9$\pm$8.3} & {2.81} & {-1.72$\pm$0.25} & {0.897$\pm$0.020}\\
        \multicolumn{1}{l}{XMP0916+5002} & {0.0497} & {206.2$\pm$4.7} & {2.48} & {-2.12$\pm$0.41} & {0.918$\pm$0.018}\\
        \multicolumn{1}{l}{XMP0916+0257} & {0.0385} & {421.2$\pm$6.0} & {1.75} & {$<$-2.03} & {0.786$\pm$0.011}\\
        \multicolumn{1}{l}{XMP0922+6324} & {0.0395} & {457.0$\pm$5.7} & {2.63} & {-2.00$\pm$0.16} & {0.728$\pm$0.014}\\
        \multicolumn{1}{l}{XMP0928+3601} & {0.0312} & {244.1$\pm$6.1} & {0.26} & {$<$-1.89} & {0.890$\pm$0.018}\\
        \multicolumn{1}{l}{XMP0930+4934} & {0.0247} & {507.0$\pm$8.5} & {3.70} & {-1.79$\pm$0.12} & {0.895$\pm$0.014}\\
        \multicolumn{1}{l}{XMP0931+2617} & {0.0638} & {164.1$\pm$5.5} & {-1.60} & {$<$-1.70} & {0.804$\pm$0.033}\\
        \multicolumn{1}{l}{XMP1003+2746} & {0.0398} & {219.1$\pm$5.9} & {0.27} & {$<$-1.73} & {0.783$\pm$0.031}\\
        \multicolumn{1}{l}{XMP1030+3151} & {0.0436} & {1617$\pm$12} & {4.74} & {-2.41$\pm$0.13} & {0.782$\pm$0.007}\\
        \multicolumn{1}{l}{XMP1032+5035} & {0.0318} & {123.9$\pm$6.2} & {1.03} & {$<$-1.29} & {0.948$\pm$0.035}\\
        \multicolumn{1}{l}{XMP1035+3814} & {0.0254} & {553.1$\pm$6.9} & {3.67} & {-1.94$\pm$0.14} & {0.900$\pm$0.014}\\
        \multicolumn{1}{l}{XMP1139+0040} & {0.0418} & {281.5$\pm$5.5} & {0.61} & {$<$-1.87} & {0.784$\pm$0.019}\\
        \multicolumn{1}{l}{XMP1140+5037} & {0.0278} & {270.8$\pm$5.7} & {2.07} & {-1.93$\pm$0.25} & {0.812$\pm$0.020}\\
        \multicolumn{1}{l}{XMP1214+1245} & {0.0192} & {154.3$\pm$6.1} & {1.00} & {$<$-1.59} & {0.314$\pm$0.031}\\
        \multicolumn{1}{l}{XMP1228+4313} & {0.0017} & {2001$\pm$21} & {9.32} & {-1.86$\pm$0.08} & {0.649$\pm$0.013}\\
        \multicolumn{1}{l}{XMP1230+0544} & {0.0397} & {421.8$\pm$9.6} & {1.58} & {$<$-1.69} & {0.869$\pm$0.019}\\
        \multicolumn{1}{l}{XMP1238+3246} & {0.0011} & {360.4$\pm$5.6} & {1.56} & {$<$-1.94} & {0.165$\pm$0.019}\\
        \multicolumn{1}{l}{XMP1322+2251} & {0.0373} & {131.6$\pm$4.5} & {0.14} & {$<$-1.55} & {0.820$\pm$0.029}\\
        \multicolumn{1}{l}{XMP1329+2237} & {0.0247} & {2246$\pm$16} & {11.65} & {-1.82$\pm$0.04} & {0.836$\pm$0.007}\\
        \multicolumn{1}{l}{XMP1344+0621} & {0.0229} & {661$\pm$11} & {2.69} & {-2.26$\pm$0.29} & {0.799$\pm$0.018}\\
        \multicolumn{1}{l}{XMP1347+0755} & {0.0438} & {848.5$\pm$8.0} & {0.65} & {$<$-2.36} & {0.844$\pm$0.009}\\
        \multicolumn{1}{l}{XMP1408+1753} & {0.0238} & {5116$\pm$23} & {13.13} & {-2.26$\pm$0.04} & {0.823$\pm$0.004}\\
        \multicolumn{1}{l}{XMP1422+5414} & {0.0212} & {2904$\pm$22} & {4.27} & {-2.52$\pm$0.18} & {0.844$\pm$0.007}\\
        \multicolumn{1}{l}{XMP1631+4426} & {0.0313} & {133.5$\pm$4.0} & {1.43} & {$<$-1.54} & {0.264$\pm$0.031}\\
        \multicolumn{1}{l}{XMP1638+2421} & {0.0344} & {630.2$\pm$8.0} & {1.10} & {$<$-2.08} & {0.928$\pm$0.013}\\
        \multicolumn{1}{l}{XMP1655+6337} & {0.0211} & {522$\pm$12} & {0.92} & {$<$-1.81} & {0.409$\pm$0.023}\\
        \multicolumn{1}{l}{XMP2048$-$0559} & {0.0480} & {719$\pm$28} & {1.15} & {$<$-1.46} & {$\ldots$}\\
        \multicolumn{1}{l}{XMP2136$-$0307} & {0.0536} & {1736.6$\pm$5.8} & {17.26} & {-2.01$\pm$0.03} & {0.779$\pm$0.003}\\
        \multicolumn{1}{l}{XMP2149$-$0535} & {0.0542} & {237$\pm$20} & {0.02} & {$<$-1.26} & {$\ldots$}\\
        \multicolumn{1}{l}{XMP2156+0856} & {0.0118} & {3506$\pm$58} & {2.76} & {-1.89$\pm$0.16} & {$\ldots$}\\
        \multicolumn{1}{l}{XMP2212+2205} & {0.0288} & {3320$\pm$50} & {3.95} & {-1.90$\pm$0.16} & {$\ldots$}\\
        \multicolumn{1}{l}{XMP2325+2008} & {0.0391} & {1992$\pm$41} & {2.46} & {-1.87$\pm$0.22} & {$\ldots$}\\
        \multicolumn{1}{l}{XMP2329+0226} & {0.0293} & {2327$\pm$45} & {2.39} & {-2.11$\pm$0.33} & {$\ldots$}\\
        \multicolumn{1}{l}{XMP2331+2226} & {0.0231} & {609$\pm$20} & {1.98} & {$<$-1.41} & {$\ldots$}\\
        \multicolumn{1}{l}{XMP2336$-$0404} & {0.0303} & {1190.6$\pm$5.1} & {4.67} & {-2.22$\pm$0.06} & {0.703$\pm$0.004} \\
        \hline
    \end{tabular}
    \end{center}
    {\raggedright
    \scriptsize 
    $^a$ The integrated flux of the H$\alpha$ line. The unit is $10^{-17}$erg~s$^{-1}$~cm$^{-2}$. \\
    $^b$ The significance of the N2 measurements. The \rNii\ and \Ha\ lines are fitted with Gaussian profiles using $\chi^2$ minimisation, where the signal represents the amplitude ratio of the fits and the noise denotes the uncertainty. \\}
\end{table*}
Using data from INT and SOAR, we have collected 45 spectroscopically confirmed XMPs, including 28 new discoveries. In this study, our primary focus is the N2 index for validating the CNN pipeline and deriving oxygen abundances via strong line diagnostics. This work provides 29 first spectra and 36 new N2 measurements. Where possible, we also measured the \rOiii/\Hb\ flux ratio (i.e. the O3 index; O3\,$\equiv\log$\{\rOiii/\Hb\}), to assess the existence of AGN activity in our samples. 
The O3 index will be used to improve our pipeline in future work. However, note that 8 XMPs observed during INT 2023B do not have the necessary coverage for O3 index measurement.


\subsection{The Fluxes and Ratios of Emission Lines}
\label{sec:line_ratio}
The fluxes of emission lines, \rNii\ and \Ha\/, are initially fit with Gaussian models using a $\chi^2$ minimisation Absorption LIne Software (ALIS; see more details in \citealt{Cooke2014}). We use this fitting method to assess whether there is a clear detection of \rNii. If the \rNii\ emission line is confidently detected ($S/N \geq 2$), we then measure the integrated fluxes of \rNii\ and \Ha\ lines by summing the pixel values associated with the target line. Conversely, if the \rNii\ line is not confidently detected ($S/N < 2$), we use the $2\sigma$ value from the Gaussian profile fitting to provide an upper limit on the N2 index.

When calculating the integrated fluxes of emission lines, we fit the continua of observed spectra with a quartic polynomial function. Since the \Ha\ and \rNii\ lines may blend together, we first integrate the flux across both lines (as a measurement of the \Ha$+$\rNii\ flux. We then refit the curve between the \Ha\ and \rNii\ lines with quartic polynomial function to obtain an accurate measurement of the \rNii\ line flux alone. By subtracting the two fluxes, we obtain the \Ha\ flux measurement for calculating the N2 index. When calculating the O3 index, the fluxes of the \rOiii\ and \Hb\ lines are integrated separately. The fitting results are compiled in Table~\ref{tab:lines}. 

For the samples with detectable \rNii\ line ($S/N\geq2$), the measurements using Gaussian fitting and integrated fluxes are consistent with each other, with a MAE of $\sim0.11$ dex. 

From the literature, there are overlapping N2 measurements for 9 objects. If the literature reported the flux ratio between \rNii+\bNii\ and \Ha\ lines, we converted their quoted (\bNii+\rNii)/\Ha\ ratio to \rNii\/Ha\ line by multiplying their line ratio by a constant of 2.96/3.96 (i.e. we assume the \rNii/\bNii =2.96; \citealt{Tachiev2001}). The comparison between literature values and our measurements ($S/N\geq2$) are consistent with each other to within a MAE of 0.086 dex.



\begin{figure}{}
\begin{center}
\graphicspath{}
        \includegraphics[width=\columnwidth]{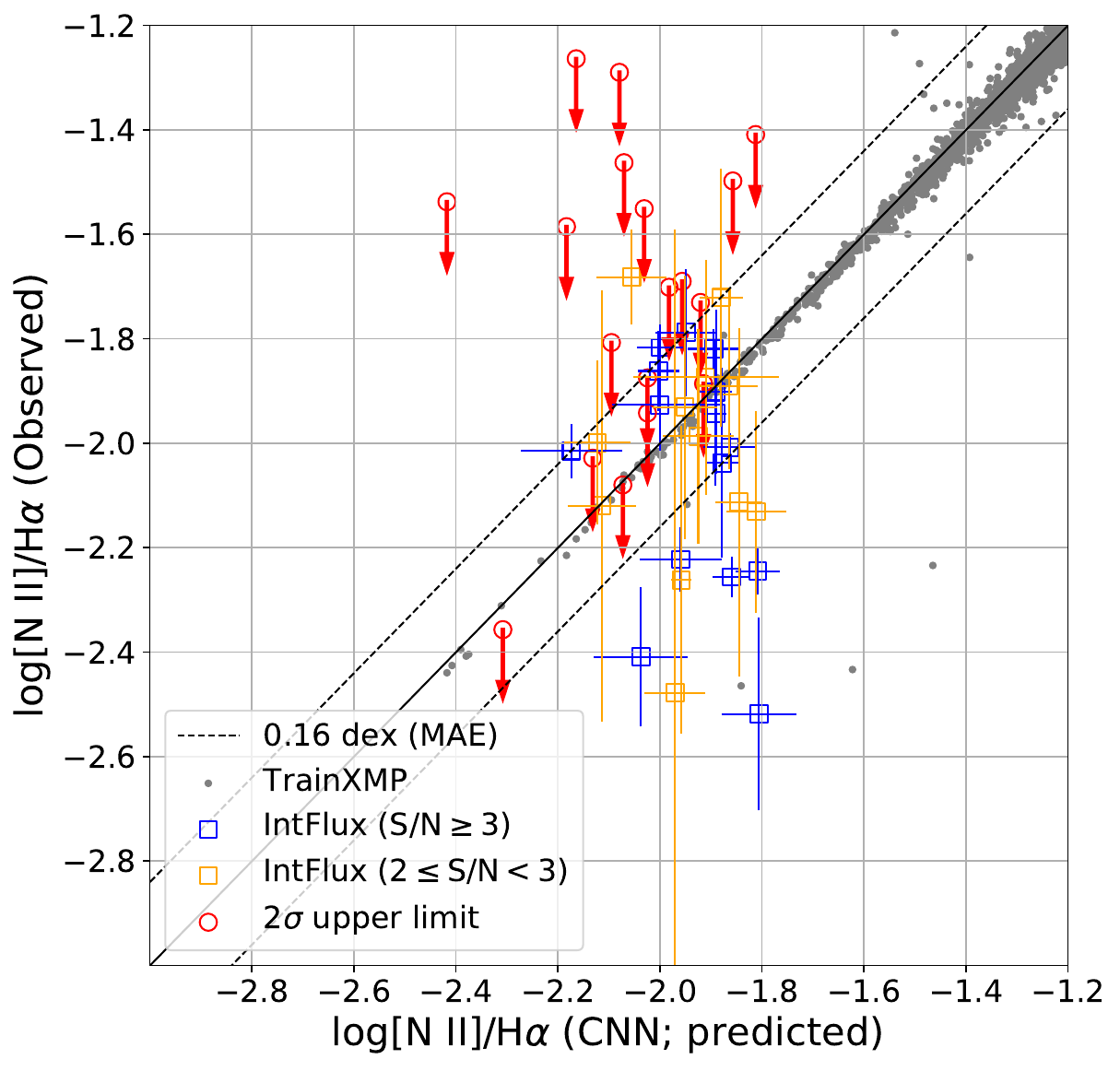}
   	\caption{Comparison of the N2 index between CNN predictions and the observed values. The gray dots show the values of our training samples. The dashed lines indicate the MAE measured using the samples marked as squares. The blue and orange squares represent the values measured using integrated flux with good ($S/N\geq3$) and fair ($2\leq{S/N}<3$) detection of \rNii\ line, respectively. The red circles are $2\sigma$ upper limits, due to the lack of a detectable \rNii\ line.}
    \label{fig:cnn2obs_N2}
\end{center}
\end{figure}

\subsection{The N2 index: CNN versus observation}

Fig.~\ref{fig:cnn2obs_N2} shows the comparison of the N2 index between our CNN predictions and our observational measurements. The MAE of the values with detectable \rNii\ line ($S/N\geq2$; squares in Fig.~\ref{fig:cnn2obs_N2}) is around 0.16 dex. We use this value to distinguish outliers. 
The outliers above the upper dashed line mostly lack robust detection of \rNii\ lines. On the other hand, we found that a subset of samples tends to have lower N2 values than those predicted by the CNN model, falling below the bottom dashed line (hereafter, lower outliers). 
We do not find any evident differences in the visual morphology and colour distributions between the lower outliers and non-outliers. This indicates that our CNN pipeline behaves correctly by assigning the values corresponding to their visual appearances. However, these galaxies somehow have lower N2 values than their visual appearances suggest. This also means that some of our new observations behave differently to the majority of our training sets. Interestingly, the outliers from our new observations occupy a parameter space similar to the three outliers in the training set (see Figure~\ref{fig:cnn2obs_N2}). While we can artificially balance the data across different ranges of N2 values, the diversity of XMPs present in the training sets is fixed. Therefore, we propose that our XMP sample shows some differences to the majority of the training set.

If this difference has a physical origin, there are two possible causes: (1) these galaxies may have lost their nitrogen gas during their evolutionary history without altering their colour or morphology; or (2) their nitrogen gas may have a distinct origin compared to most of the galaxies in our training set \citep{Chiappini2003,Pilyugin2003,Roy2021}.The first reason would require substantial outflows from these galaxies or gas stripping effects from the environments in which they reside. Investigating this possibility requires more extensive and deeper spectroscopic observations to detect any features indicating the flows, as well as understanding their environments. As for the second possibility, one could examine the N/O and O/H relationship. Secondary nitrogen production in stars occurs via the CNO cycle, catalysed by the carbon that was already present in the interstellar medium before the star was born  \citep[ISM; e.g.][]{Meynet2002}. In contrast, primary production happens when the carbon catalyst is derived directly from the helium-burning core rather than from the ISM \citep{Marigo2001}. If primary production dominates, the N/O ratio should be independent of O/H, whereas a correlation between N/O and O/H would indicate dominant secondary production. Currently, we cannot check this hypothesis because the $[\mathrm{O}\,{\textsc{ii}}]\,\lambda\lambda3727,3730$ doublet, which is necessary to determine the N/O ratio, is not covered by our observations. These possibilities will be investigated in the future with our follow-up observations.

\subsection{Derived Physical Properties}
\label{sec:physical_properties}
\begin{table*}
    \caption{Derived physical properties of the XMP galaxies reported in this work. Oxygen abundances are derived by the N2 index (Y07). For samples with significance $S/N<2$, we report a $2\sigma$ upper limit on the oxygen abundance. Calculations of luminosity distance ($D_L$), H$\alpha$ luminosities ($L\left (\mathrm{H}\alpha \right)$), star formation rate (SFR), and stellar mass ($M_*$) are described in Section~\ref{sec:physical_properties}.}
    \begin{tabular}{lcccccc} 
        \hline
        \multicolumn{1}{c}{Name} & {\OH} & {$D_L$} & {$D_L$ (corrected)} & {$L\left (\mathrm{H}\alpha \right)$} & {SFR} & {$M_*$} \\
        \multicolumn{1}{c}{} & {} & {(Mpc)} & {(Mpc)} & {($\times{10^{39}}$erg~s$^{-1}$)} & {($\times{10^{-3}}M_{\odot}~\mathrm{yr}^{-1}$)} & {($\times{10^{6}}M_{\odot}$)} \\
        \hline\hline
        \multicolumn{1}{l}{XMP0013+1354} & {7.60$\pm$0.24} & {241.99} & {245.2$\pm$1.8} & {210.2$\pm$4.7} & {922$\pm$20} & {7100$\pm$1600}\\
        \multicolumn{1}{l}{XMP0124+0838} & {7.58$\pm$0.16} & {224.39} & {228.3$\pm$1.7} & {56.91$\pm$0.89} & {249.8$\pm$3.9} & {1510$\pm$350}\\
        \multicolumn{1}{l}{XMP0219$-$0059} & {7.39$\pm$0.16} & {38.26} & {40.66$\pm$0.32} & {3.853$\pm$0.063} & {16.91$\pm$0.27} & {0.57$\pm$0.13}\\
        \multicolumn{1}{l}{XMP0429+0026} & {7.56$\pm$0.22} & {53.37} & {51.93$\pm$0.62} & {1.675$\pm$0.048} & {7.35$\pm$0.21} & {1.19$\pm$0.28}\\
        \multicolumn{1}{l}{XMP0742+1103} & {$<$8.47} & {201.47} & {200.7$\pm$1.5} & {6.58$\pm$0.23} & {28.9$\pm$1.0} & {198$\pm$46}\\
        \multicolumn{1}{l}{XMP0752+2340} & {7.65$\pm$0.17} & {218.53} & {221.5$\pm$1.7} & {61.5$\pm$1.1} & {270.0$\pm$4.9} & {2990$\pm$690}\\
        \multicolumn{1}{l}{XMP0801+2640} & {7.74$\pm$0.16} & {120.11} & {123.36$\pm$0.94} & {72.2$\pm$1.2} & {316.7$\pm$5.2} & {120$\pm$28}\\
        \multicolumn{1}{l}{XMP0803+1635} & {7.48$\pm$0.23} & {95.48} & {100.8$\pm$1.6} & {5.36$\pm$0.19} & {23.51$\pm$0.83} & {39.6$\pm$9.2}\\
        \multicolumn{1}{l}{XMP0827+1059} & {7.19$\pm$0.76} & {200.26} & {203.7$\pm$1.5} & {16.25$\pm$0.45} & {71.3$\pm$2.0} & {500$\pm$110}\\
        \multicolumn{1}{l}{XMP0850+1150} & {7.86$\pm$0.18} & {133.03} & {133.2$\pm$1.1} & {13.17$\pm$0.29} & {57.8$\pm$1.3} & {86$\pm$20}\\
        \multicolumn{1}{l}{XMP0856+2414} & {7.82$\pm$0.26} & {235.91} & {237.2$\pm$1.8} & {19.91$\pm$0.63} & {87.4$\pm$2.8} & {1170$\pm$270}\\
        \multicolumn{1}{l}{XMP0916+5002} & {7.49$\pm$0.38} & {229.22} & {232.4$\pm$1.7} & {13.32$\pm$0.36} & {58.5$\pm$1.6} & {403$\pm$93}\\
        \multicolumn{1}{l}{XMP0916+0257} & {$<$8.07} & {176.27} & {176.5$\pm$1.7} & {15.70$\pm$0.37} & {68.9$\pm$1.6} & {157$\pm$36}\\
        \multicolumn{1}{l}{XMP0922+6324} & {7.59$\pm$0.21} & {180.99} & {180.4$\pm$1.4} & {17.79$\pm$0.35} & {78.1$\pm$1.5} & {125$\pm$29}\\
        \multicolumn{1}{l}{XMP0928+3601} & {$<$8.17} & {142.17} & {145.2$\pm$1.2} & {6.16$\pm$0.18} & {27.03$\pm$0.81} & {25.4$\pm$5.9}\\
        \multicolumn{1}{l}{XMP0930+4934} & {7.77$\pm$0.19} & {111.97} & {106.41$\pm$0.89} & {6.87$\pm$0.16} & {30.15$\pm$0.71} & {21.1$\pm$4.9}\\
        \multicolumn{1}{l}{XMP0931+2617} & {$<$8.28} & {297.41} & {297.2$\pm$3.5} & {17.34$\pm$0.71} & {76.1$\pm$3.1} & {1500$\pm$350}\\
        \multicolumn{1}{l}{XMP1003+2746} & {$<$8.27} & {182.32} & {187.2$\pm$1.4} & {9.19$\pm$0.29} & {40.3$\pm$1.3} & {164$\pm$38}\\
        \multicolumn{1}{l}{XMP1030+3151} & {7.25$\pm$0.19} & {200.11} & {203.5$\pm$1.5} & {80.2$\pm$1.4} & {351.8$\pm$6.0} & {840$\pm$190}\\
        \multicolumn{1}{l}{XMP1032+5035} & {$<$8.65} & {145.04} & {151.9$\pm$1.2} & {3.42$\pm$0.18} & {15.02$\pm$0.79} & {324$\pm$75}\\
        \multicolumn{1}{l}{XMP1035+3814} & {7.64$\pm$0.20} & {115.08} & {118.27$\pm$0.88} & {9.26$\pm$0.18} & {40.63$\pm$0.79} & {30.8$\pm$7.1}\\
        \multicolumn{1}{l}{XMP1139+0040} & {$<$8.16} & {191.86} & {186.5$\pm$1.4} & {11.72$\pm$0.29} & {51.4$\pm$1.3} & {301$\pm$69}\\
        \multicolumn{1}{l}{XMP1140+5037} & {7.65$\pm$0.26} & {126.21} & {122.38$\pm$0.98} & {4.85$\pm$0.13} & {21.30$\pm$0.56} & {78$\pm$18}\\
        \multicolumn{1}{l}{XMP1214+1245} & {$<$8.44} & {86.82} & {88.07$\pm$0.68} & {1.432$\pm$0.061} & {6.28$\pm$0.27} & {70$\pm$16}\\
        \multicolumn{1}{l}{XMP1228+4313} & {7.71$\pm$0.17} & {7.69} & {4.87$\pm$0.13} & {0.0568$\pm$0.0030} & {0.249$\pm$0.013} & {0.0251$\pm$0.0059}\\
        \multicolumn{1}{l}{XMP1230+0544} & {$<$8.33} & {181.72} & {183.4$\pm$1.4} & {16.98$\pm$0.46} & {74.5$\pm$2.0} & {325$\pm$75}\\
        \multicolumn{1}{l}{XMP1238+3246} & {$<$8.19} & {4.87} & {8.14$\pm$0.66} & {0.0286$\pm$0.0046} & {0.125$\pm$0.020} & {0.0196$\pm$0.0055}\\
        \multicolumn{1}{l}{XMP1322+2251} & {$<$8.41} & {170.74} & {171.2$\pm$1.3} & {4.62$\pm$0.17} & {20.26$\pm$0.75} & {74$\pm$17}\\
        \multicolumn{1}{l}{XMP1329+2237} & {7.75$\pm$0.16} & {111.80} & {113.95$\pm$0.87} & {34.89$\pm$0.59} & {153.1$\pm$2.6} & {151$\pm$35}\\
        \multicolumn{1}{l}{XMP1344+0621} & {7.37$\pm$0.29} & {103.60} & {103.1$\pm$3.1} & {8.41$\pm$0.53} & {36.9$\pm$2.3} & {18.8$\pm$4.5}\\
        \multicolumn{1}{l}{XMP1347+0755} & {$<$7.77} & {201.43} & {201.4$\pm$1.5} & {41.19$\pm$0.72} & {180.8$\pm$3.2} & {640$\pm$150}\\
        \multicolumn{1}{l}{XMP1408+1753} & {7.38$\pm$0.16} & {107.82} & {109.86$\pm$0.97} & {73.9$\pm$1.3} & {324.3$\pm$5.9} & {14.2$\pm$3.3}\\
        \multicolumn{1}{l}{XMP1422+5414} & {7.16$\pm$0.22} & {95.79} & {98.45$\pm$0.73} & {33.68$\pm$0.56} & {147.8$\pm$2.5} & {38.0$\pm$8.8}\\
        \multicolumn{1}{l}{XMP1631+4426} & {$<$8.43} & {142.33} & {146.4$\pm$1.1} & {3.42$\pm$0.11} & {15.03$\pm$0.50} & {16.5$\pm$3.8}\\
        \multicolumn{1}{l}{XMP1638+2421} & {$<$7.98} & {156.82} & {159.6$\pm$1.2} & {19.20$\pm$0.39} & {84.3$\pm$1.7} & {82$\pm$19}\\
        \multicolumn{1}{l}{XMP1655+6337} & {$<$8.21} & {95.46} & {100.88$\pm$0.75} & {6.36$\pm$0.17} & {27.90$\pm$0.75} & {5.2$\pm$1.2}\\
        \multicolumn{1}{l}{XMP2048$-$0559} & {$<$8.49} & {220.98} & {217.7$\pm$1.8} & {40.8$\pm$1.7} & {179.0$\pm$7.6} & {1190$\pm$270}\\
        \multicolumn{1}{l}{XMP2136$-$0307} & {7.58$\pm$0.16} & {248.06} & {247.8$\pm$1.8} & {127.6$\pm$1.9} & {559.9$\pm$8.5} & {4010$\pm$930}\\
        \multicolumn{1}{l}{XMP2149$-$0535} & {$<$8.64} & {251.08} & {253.0$\pm$1.9} & {18.1$\pm$1.5} & {79.6$\pm$6.7} & {1280$\pm$290}\\
        \multicolumn{1}{l}{XMP2156+0856} & {7.68$\pm$0.21} & {52.83} & {51.94$\pm$0.39} & {11.31$\pm$0.25} & {49.7$\pm$1.1} & {3.47$\pm$0.80}\\
        \multicolumn{1}{l}{XMP2212+2205} & {7.67$\pm$0.21} & {130.70} & {133.74$\pm$1.00} & {71.1$\pm$1.5} & {311.8$\pm$6.6} & {262$\pm$61}\\
        \multicolumn{1}{l}{XMP2325+2008} & {7.70$\pm$0.25} & {179.25} & {174.6$\pm$1.4} & {72.7$\pm$1.9} & {318.9$\pm$8.3} & {1370$\pm$320}\\
        \multicolumn{1}{l}{XMP2329+0226} & {7.50$\pm$0.32} & {133.18} & {134.65$\pm$0.99} & {50.5$\pm$1.2} & {221.6$\pm$5.4} & {369$\pm$85}\\
        \multicolumn{1}{l}{XMP2331+2226} & {$<$8.55} & {104.64} & {97.8$\pm$1.2} & {6.97$\pm$0.28} & {30.6$\pm$1.2} & {26.5$\pm$6.1}\\
        \multicolumn{1}{l}{XMP2336$-$0404} & {7.40$\pm$0.17} & {137.63} & {138.8$\pm$1.1} & {27.46$\pm$0.44} & {120.5$\pm$1.9} & {94$\pm$22}\\
        \hline
    \end{tabular}
    \label{tab:properties}
\end{table*}

We have derived several physical properties of the XMP galaxies observed with our programme, including the oxygen abundance, star formation rate, and stellar mass; these values are summarised in Table~\ref{tab:properties}.

\subsubsection{Oxygen Abundance}
\begin{figure*}{}
\begin{center}
\graphicspath{}
        \includegraphics[width=2\columnwidth]{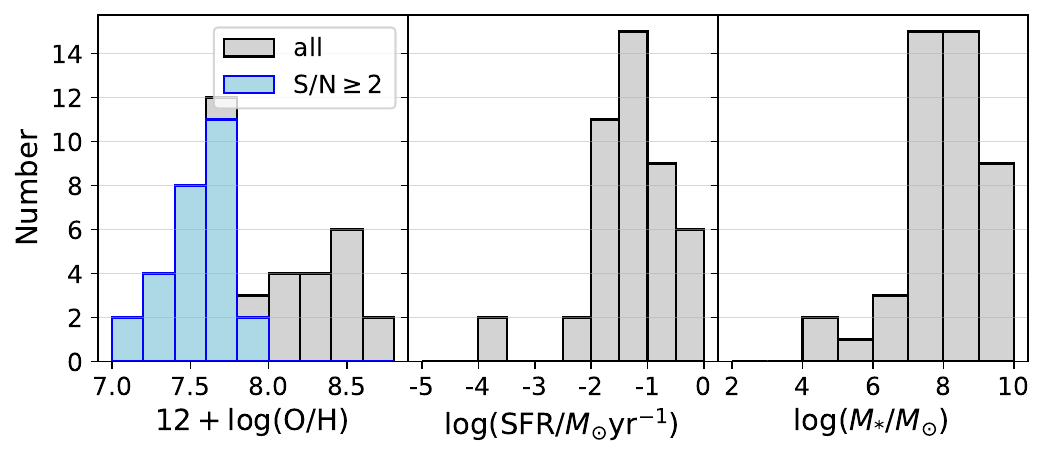}
   	\caption{The distributions of oxygen abundance, star formation rate, and stellar mass of our XMP galaxy sample. The gray histograms show the values of all samples, while the blue histograms in the first panel represent the values of the samples with significantly detected \rNii\ lines ($S/N\geq2$).}
    \label{fig:physprop}
\end{center}
\end{figure*}
The oxygen abundance is estimated using the observed N2 values together with the empirical relation from Y07: 
\begin{equation}
    12+\log \left ( \mathrm{O/H} \right )=9.263+0.836\times{N2}.
\end{equation}
The relation is based on a linear least-squares fit to the data collected from SDSS and various literature sources, providing a better description of the data compared to the empirical relation of \citet{Pettini2004}. The uncertainty of this linear fit to the data used in Y07 is 0.159 dex. If a galaxy's \rNii\ line is not detected, we report a $2\sigma$ upper limit on the oxygen abundance. The estimated oxygen abundance of our samples ranges between $7.1\leq$\OH$\leq8.7$ ($2\sigma$ upper limit). We have 21 samples with estimated oxygen abundances of \OH$\leq7.7$ ($\sim0.1~Z_{\odot}$), and 18 samples are reported with $2\sigma$ upper limits. Planned follow-up observations of these targets will firmly pin down the chemistry of these near-pristine galaxies. The distribution of oxygen abundance estimates is shown in the leftmost panel of Fig.~\ref{fig:physprop}. 

\subsubsection{Distance, H$\alpha$ luminosity and star formation rate}
For distance-derived properties, such as luminosity and star formation rate, we use the luminosity distance to estimate these quantities. The luminosity distance ($D_L$) is calculated using spectroscopic redshifts from our observations in combination with \textsc{Astropy}'s cosmology package \citep{Astropy2022}. We assume a flat $\Lambda$CDM cosmology with $H_0=67.4\pm0.5~\mathrm{km}~\mathrm{s}^{-1}\mathrm{Mpc}^{-1}$ and $\Omega_{m}=0.315\pm0.007$ \citep{Planck2020}. The distance is corrected for peculiar velocity based on the results from \citet{Carrick2015}. Both the original ($D_L$) and corrected [$D_L$ (corrected)] values are listed in Table~\ref{tab:properties}. The quoted uncertainties include the uncertainties on the redshift, peculiar velocity, and cosmological parameters.
The H$\alpha$ luminosity, $L\left (\mathrm{H}\alpha \right)$, is then calculated by the following conversion using the corrected luminosity distances and the measured H$\alpha$ fluxes: 
\begin{equation}
    L\left (\mathrm{H}\alpha \right) = F\left (\mathrm{H}\alpha \right)4\pi{D_L}^2.
\end{equation} 
The star formation rate (SFR) can be estimated using hydrogen emission lines such as \Ha\ line. This line is produced by the recombination of ionised hydrogen in the H\textsc{ii} regions. Thus, the H$\alpha$ luminosity is linked with the number of ionising photons and traces the formation of young ($<20$Myr), massive ($>10M_{\odot}$) stars. The following conversion between SFR and H$\alpha$ luminosity is provided in \citet{Kennicutt1998} derived by \citet{Kennicutt1994} and \citet{Madau1998}. 
\begin{equation}
    \mathrm{SFR} = 7.9\times10^{-42}~\times(L\left (\mathrm{H}\alpha \right)/{\rm erg~s}^{-1})~M_{\odot}~{\rm yr}^{-1}. 
\end{equation}
Note that this relationship assumes a Salpeter initial mass function \citep[IMF;][]{Salpeter1955}. We apply a correction factor of 1.8 to convert this relationship to a \citet{Chabrier2003} IMF. The resulting values of $\log_{10}\mathrm{(SFR)}$ are between -3.9 and -0.035, with the median value of $-1.16\pm0.60$, which includes $\sim$69\% of the samples. The SFR distribution of our sample is shown in the middle panel of Fig.~\ref{fig:physprop}.

\subsubsection{Stellar mass}
The stellar mass is estimated using stellar mass-to-light ratio, obtained with the broadband luminosity in SDSS $i$-band, in combination with the $r-i$ colour \citep[][B03]{Bell2003}. The conversion of the solar absolute magnitude to SDSS $i$-band filter is from \citet{Blanton2007}. The filter choices are to reduce contamination from strong emission lines. However, we note that this kind of relationship is not robust for metal-poor dwarf galaxies. The results presented in this work just serve as an indicative measure of the stellar mass of the XMP galaxies in our sample. We adopt the following relationship:
\begin{equation}
\label{eq:Mstar}
    \log_{10}(\frac{M_*}{L})= 0.006 + 1.114\times\left ( r-i \right ).
\end{equation}
where the value of $M_*/L$ is expressed in solar units. Another relationship, using more complex stellar models, was introduced in \citet[][Z09]{Zibetti2009}. To assess which of these aforementioned stellar mass models are more suitable for the XMP galaxies of our sample, we compared the stellar mass of Leo P estimated using both B03 and Z09 relations to the robust stellar mass measurement derived from HST imaging by \citet{McQuinn2015}. Among the two, the B03 relation yielded a closer estimate to the robust value. Therefore, for stellar mass estimation in this work, we adopt Equation~\ref{eq:Mstar} from B03, and divided the value by a factor of correct the `diet' Salpeter IMF used in B03 to \citet{Chabrier2003} IMF. The resulting values of $\log_{10}$($M_{*}/M_{\odot}$) are between 4.3 and 9.8, with the median value of $8.1\pm1.0$ including about 71\% of the sample. 
The distribution is shown in the rightmost panel of Fig.~\ref{fig:physprop}.

\subsection{AGN activity}
\label{sec:agn}
\begin{figure}{}
\begin{center}
\graphicspath{}
        \includegraphics[width=\columnwidth]{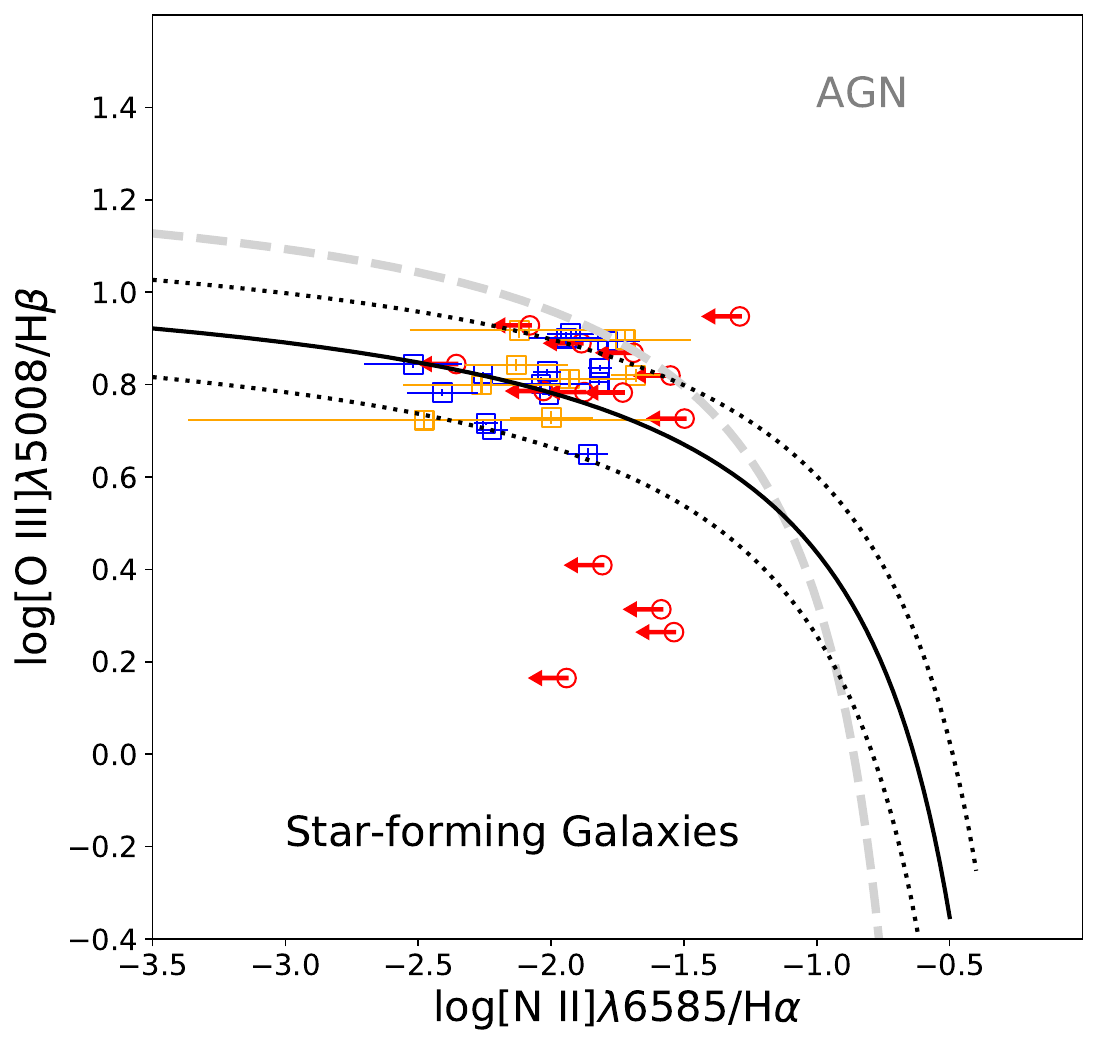}
   	\caption{Diagnostic diagram of \rNii/\Ha\ versus \rOiii/\Hb. The gray dashed line shows Equation 1 from \citet{Kauffmann2003b}, which delineates star-forming galaxies and AGN. The black solid line represents the mean of local star-forming sequences for SDSS galaxies analysed in \citet{Kewley2006}, while the dotted black lines show the error of 0.1 dex from their models \citep{Kewley2013}. The symbols show our sample of XMP galaxies, with the same colour-coding as used in Figure~\ref{fig:cnn2obs_N2}.}
    \label{fig:bpt}
\end{center}
\end{figure}

Figure~\ref{fig:bpt} shows the BPT diagram \citep{Baldwin1981} -- \rNii/\Ha\ versus \rOiii/\Hb\ diagnostic diagram -- for assessing AGN activity in our sample of XMP galaxies. The gray dashed line represents the relation provided in Equation 1 of \citet{Kauffmann2003b} to separate the population of star-forming galaxies and AGN. This indicates that our XMP galaxies do not show any indication of AGN activity. The black solid line and dotted lines show the mean of SDSS star-forming galaxies at redshift $0.04<z<0.1$ \citep{Kewley2006,Kewley2013}. This range contains 91\% of the SDSS star-forming galaxies from \citet{Kewley2006}. Four of our objects are outliers to this trend with lower O3 values. This could be due to their low metallicity nature. \citet{Groves2006} studied the evolution of diagnostic diagrams for low-metallicity AGN (lower than 1 $Z_{\odot}$). At metallicity of $0.1Z_\odot$, their simulations predicted a decrease in the \rOiii\//\Hb\ ratio (lower than 0.5). This may explain the outliers' behaviour on the diagnostic diagram, indicating that they may contain low-metallicity AGN. To test this possibility, future observations will target the [O\,\textsc{ii}], [Ne\,\textsc{iii}], and [Ne\,\textsc{v}] emission lines.

\subsection{Colour Distribution}
\label{sec:colour}
\begin{figure*}{}
\begin{center}
\graphicspath{}
	\includegraphics[width=2.1\columnwidth]{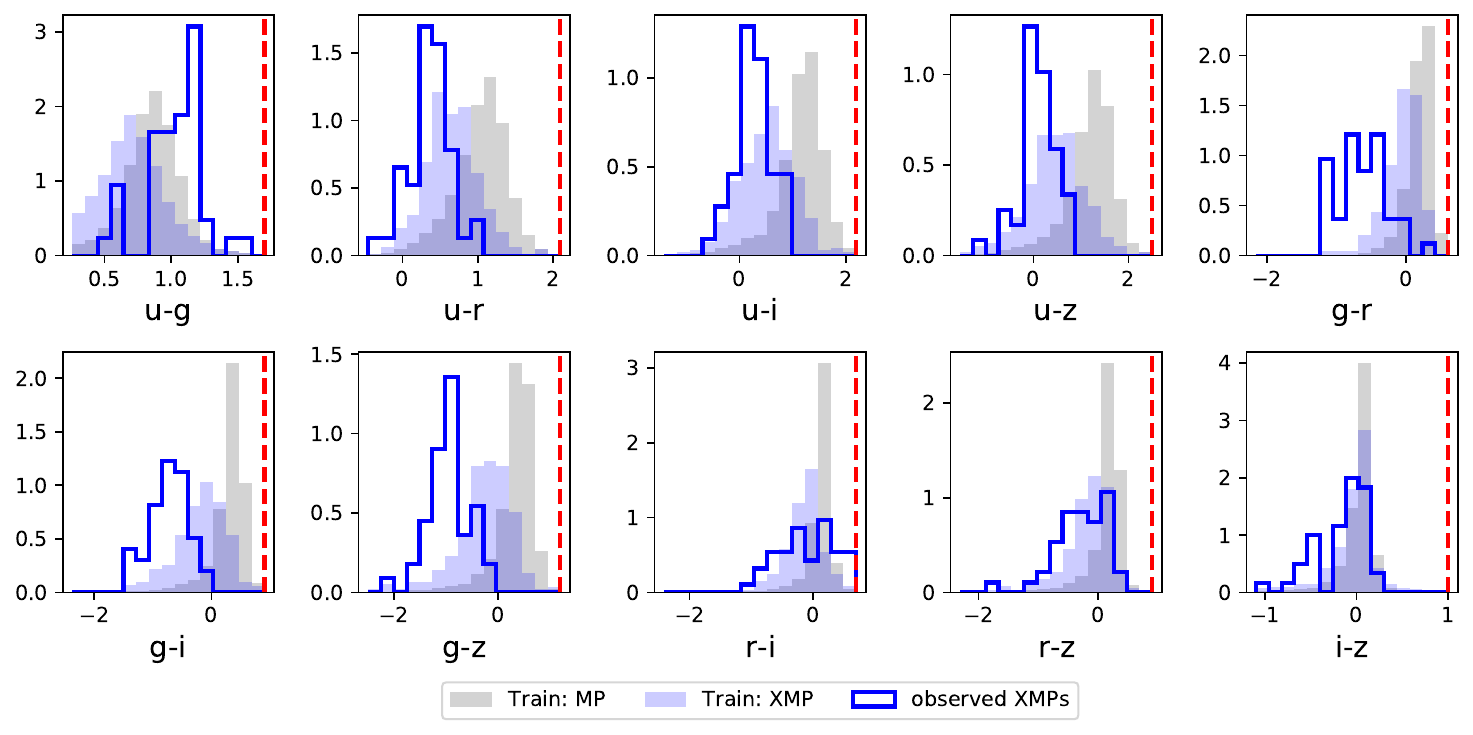}
   	\caption{Colour distributions of training MP samples (gray shadings), training XMP samples (blue shadings), and the observed XMPs (unfilled blue histogram). 
    The red dashed lines indicate the upper limit applied to each colour for querying the working samples (see Table~\ref{tab:selection_crit} and the discussion in Section~\ref{sec:target_samples}).}
    \label{fig:colour}
\end{center}
\end{figure*}
\begin{figure}{}
\begin{center}
\graphicspath{}
	\includegraphics[width=\columnwidth]{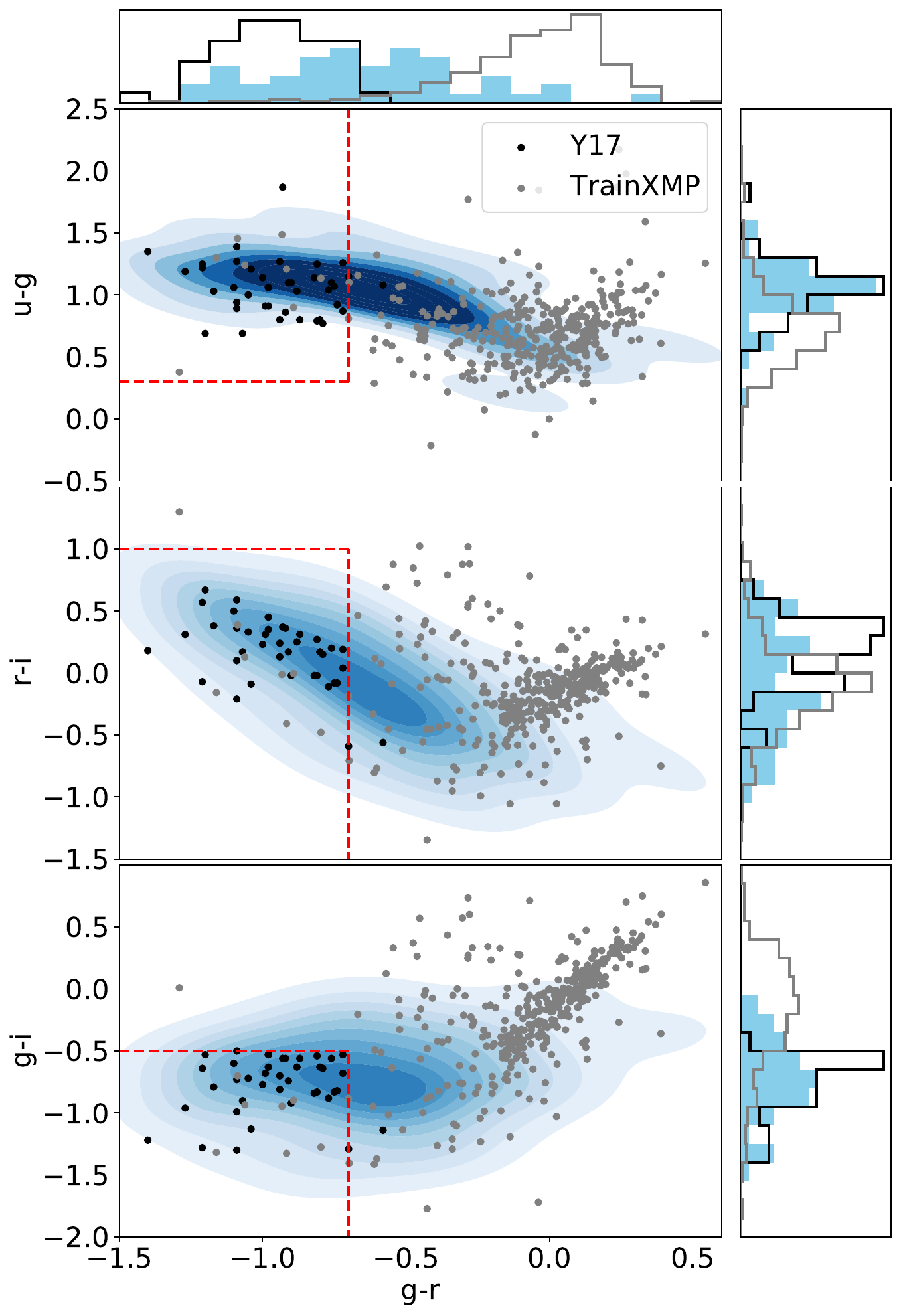}
   	\caption{The comparison of the colour-colour diagram and histograms between the observed XMP (blue contours and histograms), samples of BB galaxies from Y17 (black dots and histogram), and training XMP samples (TrainXMP) from this work (grey dots and histogram). The red dashed lines are the colour criteria used in Y17 to select green pea galaxies at $z\lessapprox0.05$.}
    \label{fig:bb_allprep}
\end{center}
\end{figure}
In Fig.~\ref{fig:colour}, we compare the colour distributions of the observed galaxies and the training samples (MP and XMP).
Note that the majority of the training XMP samples have an N2 value between -1.5 and -1.8 ($>77$\%). Thus, we expect there to be a slight difference in the colour distributions between the training XMP samples and our observed XMP galaxies with predicted $\mathrm{N2}<-1.8$. 

However, a notable discrepancy exists in the colour distributions of those calculated particularly using the $g$-band between the training samples and the observed XMPs. These samples tend to be brighter in $g$-band compared to the majority of training XMP samples. The brighter $g$-band magnitudes are likely due to significant emission from the \Oiii\ doublet, which dominates the flux in the SDSS $g$-band given the redshifts of our observed XMPs. This characteristic is similar to that of blueberry galaxies \citep[BB;][hereafter Y17]{Yang2017}, which are low-redshift counterparts of green pea galaxies \citep{Cardamone2009} and high redshift Ly$\alpha$ emitting galaxies. 

Fig.~\ref{fig:bb_allprep} shows the colour distributions of the observed XMP (blue contours), BB galaxies from Y17 (black dots), and the training XMP samples (grey dots) in this work. The red dashed lines in each panel represent the colour criteria used in Y17 to select green pea galaxies at $z\lessapprox0.05$. This figure demonstrates that the observed XMPs are mostly greener\footnote{We describe our samples as `greener' rather than `bluer' because only $g-$band appears brighter, while the $u-$band does not exhibit a similar increase in brightness.} than the majority of the training samples. About 38\% (17 XMPs) of the observed objects fall into the colour regions of BB galaxies defined by Y17. We found that the N2 values of the training samples satisfying the colour criteria of BB galaxies are lower than those outside these criteria. Specifically, the median N2 value of BB-like training XMP samples (within the red-dashed lines in Fig.~\ref{fig:bb_allprep}) is $-2.02$, while those outside these regions have a median N2 value of $-1.64$. This indicates that the phenomenon shown here are simply due to the blind selection of samples with lower N2 index.

\section{Summary}
\label{sec:summary}
To advance the discovery of new XMP galaxies, we promote the use of convolutional neural networks (CNNs) for the first time to accelerate the process of XMP identification and characterisation for current and upcoming wide-area multi-band sky surveys. A primary advantage of this approach is to efficiently consider both morphology and colour information simultaneously from broadband images.
Our DL pipeline is built from three individual CNN procedures: (i) MP classifier, (ii) XMP classifier, and (iii) N2 predictor, to conduct sequential classification and predictions of the N2 index (N2\,$\equiv\log$\{\rNii/\Ha\}) for MP galaxies. The N2 index is then used to select the most promising XMP galaxies. This design is to ensure an effective and efficient training and identification of the extremely sparse population of XMPs. Each CNN procedure contains 9 CNN models (i.e. 3 different initialisations $\times$ 3 training datasets) to account for the variation in each training run and the impact of quality of the selected training subsets. The median values of these 9 CNNs are used for each classifier and the N2 predictor.  


The trained DL pipeline is applied to the multi-band imaging data without spectroscopy from the SDSS DR17. There are 232\,954 XMP candidates selected with the criteria of $P_\mathrm{MP}>0.5$ and $P_\mathrm{XMP}>0.5$ from over 7 million SDSS galaxies. For observational candidates, we further select 390 promising candidates with $P_\mathrm{MP}>0.99$, $P_\mathrm{XMP}>0.99$, and $\mathrm{N2}<-1.8$. Among them, we successfully observed 45 XMP candidates with redshifts less than 0.065 using the 2.54~m Isaac Newton Telescope (INT) and the 4.1~m Southern Astrophysical Research (SOAR) Telescope between 2023 and 2024. All 45 observed XMP candidates are spectroscopically confirmed to be metal-poor, including 28 new discoveries. Additionally, our observations provide the first N2 measurements for 36 XMPs. These N2 measurements are found to be consistent with the CNN predictions with a MAE of 0.16 dex. However, we found a set of samples with differences between predicted and observed N2 values greater than 0.16. These objects do not show distinct morphologies and colours from those within the MAE threshold, indicating that these objects somehow possess lower N2 values than their morphologies and colours suggest. These galaxies may have experienced significant outflows or gas stripping in their evolutionary history. Another hypothesis is that their nitrogen gas may have originated from a different route compared to the majority of our training XMPs. However, as this work aims to report a new methodology and the discovery of new XMP galaxies, our observations do not have sufficient wavelength coverage and depth to validate this hypothesis. We will address this with forthcoming observations. 

The reported samples have estimated oxygen abundances of $7.1\leq$\OH$\leq8.7$ ($2\sigma$ upper limit), based on the N2 index. There are 21/45 galaxies with estimated oxygen abundances below 7.7, and 18/45 galaxies lack of detectable \rNii\ lines ($S/N<2$). These samples offer an exciting opportunity to identify a record-breaking XMP, providing valuable insights into chemical abundances and evolutionary processes within galaxies in extremely metal-deficient environments. The star formation rates of our XMPs are between $10^{-3.9}-10^{-0.035}~M_\odot/\mathrm{yr}$, and their stellar masses are in the range $10^{4.3}-10^{9.8}~M_\odot$ based on the \citet{Bell2003} calibration, with a correction to the Chabrier IMF. The BPT diagram of our XMPs shows that our objects are mostly star-forming galaxies without AGN activity. However, 4 XMPs without detectable \rNii\ lines deviate from the typical trend of star-forming galaxies on the BPT diagram, suggesting that they may potentially be low-metallicity AGN at metallicity $\lesssim0.1Z_{\odot}$. Future observations that cover the emission lines [O\,\textsc{ii}], [Ne\,\textsc{iii}], and [Ne\,\textsc{v}] at wavelengths $<4000$\,\AA\ are required to test this possibility. 

Finally, we examined the colour distributions of our observed galaxies, and found that they tend to be brighter in the SDSS $g-$band than the training samples. This is likely to be caused by significant emission of the [O\,\textsc{iii}]\,$\lambda\lambda4960, 5008$ doublet that coincide with the bandpass of this filter. This characteristic is reminiscent of green pea galaxies and high-redshift Ly$\alpha$ emitters, but at lower redshifts ($z\lessapprox0.05$), similar to blueberry galaxies. 
By applying the Y17 blueberry colour criteria, 38\% (17 XMPs) of the observed samples are categorised as blueberry galaxies. We found that our training samples, which share similar colour characteristics to the blueberry galaxies, tend to have lower N2 values. This leads to a skew of our observed samples, selected based on low predicted N2 values, towards the colour regions associated with blueberry galaxies. 

In this work, we developed a DL pipeline and validated its effectiveness using new observations. In the near-future, we will conduct followup spectroscopic observations that will cover additional key emission lines to: (1) enable direct measurements of oxygen abundances; (2) measure the primordial $^4$He abundance; and (3) address questions that were raised in our analysis, including: (a) the origin of outliers in our pipeline --- potentially due to outflows, environmental impact, or nitrogen produced through different channels; (b) test if the 4 galaxies with low \rOiii/\Hb\ and low \rNii/\Ha\ contain a low-metallicity AGN; and (c) develop the connection between our discoveries, and high-redshift galaxies.

\section*{Acknowledgements}

During this work, RJC was funded by a Royal Society University Research Fellowship. TYC and RJC acknowledge support from STFC (ST/T000244/1). The Issac Newton Telescope is operated on the island of La Palma by the Isaac Newton Group of Telescopes in the Spanish Observatorio del Roque de los Muchachos of the Instituto de Astrofísica de Canarias. The IDS spectroscopy was obtained as part of P4 programme. Based on observations obtained at the Southern Astrophysical Research (SOAR) telescope, which is a joint project of the Minist\'{e}rio da Ci\^{e}ncia, Tecnologia e Inova\c{c}\~{o}es (MCTI/LNA) do Brasil, the US National Science Foundation’s NOIRLab, the University of North Carolina at Chapel Hill (UNC), and Michigan State University (MSU).


\section*{Data Availability}
The observational data via open-time programme can be shared upon request. The catalogues for the measured quantities presented in this work will be available upon the publication of this manuscript.

\bibliographystyle{mnras}
\bibliography{ms} 

\begin{thebibliography}{}
\makeatletter
\relax
\def\mn@urlcharsother{\let\do\@makeother \do\$\do\&\do\#\do\^\do\_\do\%\do\~}
\def\mn@doi{\begingroup\mn@urlcharsother \@ifnextchar [ {\mn@doi@} {\mn@doi@[]}}
\def\mn@doi@[#1]#2{\def\@tempa{#1}\ifx\@tempa\@empty \href {http://dx.doi.org/#2} {doi:#2}\else \href {http://dx.doi.org/#2} {#1}\fi \endgroup}
\def\mn@eprint#1#2{\mn@eprint@#1:#2::\@nil}
\def\mn@eprint@arXiv#1{\href {http://arxiv.org/abs/#1} {{\tt arXiv:#1}}}
\def\mn@eprint@dblp#1{\href {http://dblp.uni-trier.de/rec/bibtex/#1.xml} {dblp:#1}}
\def\mn@eprint@#1:#2:#3:#4\@nil{\def\@tempa {#1}\def\@tempb {#2}\def\@tempc {#3}\ifx \@tempc \@empty \let \@tempc \@tempb \let \@tempb \@tempa \fi \ifx \@tempb \@empty \def\@tempb {arXiv}\fi \@ifundefined {mn@eprint@\@tempb}{\@tempb:\@tempc}{\expandafter \expandafter \csname mn@eprint@\@tempb\endcsname \expandafter{\@tempc}}}

\bibitem[\protect\citeauthoryear{{Abdurro'uf} et~al.,}{{Abdurro'uf} et~al.}{2022}]{sdss17-2022}
{Abdurro'uf} et~al., 2022, \mn@doi [\apjs] {10.3847/1538-4365/ac4414}, \href {https://ui.adsabs.harvard.edu/abs/2022ApJS..259...35A} {259, 35}

\bibitem[\protect\citeauthoryear{{Ann}, {Seo}  \& {Ha}}{{Ann} et~al.}{2015}]{Ann2015}
{Ann} H.~B.,  {Seo} M.,   {Ha} D.~K.,  2015, \mn@doi [\apjs] {10.1088/0067-0049/217/2/27}, \href {https://ui.adsabs.harvard.edu/abs/2015ApJS..217...27A} {217, 27}

\bibitem[\protect\citeauthoryear{{Astropy Collaboration} et~al.,}{{Astropy Collaboration} et~al.}{2022}]{Astropy2022}
{Astropy Collaboration} et~al., 2022, \mn@doi [\apj] {10.3847/1538-4357/ac7c74}, \href {https://ui.adsabs.harvard.edu/abs/2022ApJ...935..167A} {935, 167}

\bibitem[\protect\citeauthoryear{{Aver}, {Berg}, {Olive}, {Pogge}, {Salzer}  \& {Skillman}}{{Aver} et~al.}{2021}]{Aver2021}
{Aver} E.,  {Berg} D.~A.,  {Olive} K.~A.,  {Pogge} R.~W.,  {Salzer} J.~J.,   {Skillman} E.~D.,  2021, \mn@doi [\jcap] {10.1088/1475-7516/2021/03/027}, \href {https://ui.adsabs.harvard.edu/abs/2021JCAP...03..027A} {2021, 027}

\bibitem[\protect\citeauthoryear{{Baldwin}, {Phillips}  \& {Terlevich}}{{Baldwin} et~al.}{1981}]{Baldwin1981}
{Baldwin} J.~A.,  {Phillips} M.~M.,   {Terlevich} R.,  1981, \mn@doi [\pasp] {10.1086/130766}, \href {https://ui.adsabs.harvard.edu/abs/1981PASP...93....5B} {93, 5}

\bibitem[\protect\citeauthoryear{{Bell}, {McIntosh}, {Katz}  \& {Weinberg}}{{Bell} et~al.}{2003}]{Bell2003}
{Bell} E.~F.,  {McIntosh} D.~H.,  {Katz} N.,   {Weinberg} M.~D.,  2003, \mn@doi [\apjs] {10.1086/378847}, \href {https://ui.adsabs.harvard.edu/abs/2003ApJS..149..289B} {149, 289}

\bibitem[\protect\citeauthoryear{{Blanton} \& {Roweis}}{{Blanton} \& {Roweis}}{2007}]{Blanton2007}
{Blanton} M.~R.,  {Roweis} S.,  2007, \mn@doi [\aj] {10.1086/510127}, \href {https://ui.adsabs.harvard.edu/abs/2007AJ....133..734B} {133, 734}

\bibitem[\protect\citeauthoryear{{Brinchmann}, {Charlot}, {White}, {Tremonti}, {Kauffmann}, {Heckman}  \& {Brinkmann}}{{Brinchmann} et~al.}{2004}]{Brinchmann2004}
{Brinchmann} J.,  {Charlot} S.,  {White} S.~D.~M.,  {Tremonti} C.,  {Kauffmann} G.,  {Heckman} T.,   {Brinkmann} J.,  2004, \mn@doi [\mnras] {10.1111/j.1365-2966.2004.07881.x}, \href {https://ui.adsabs.harvard.edu/abs/2004MNRAS.351.1151B} {351, 1151}

\bibitem[\protect\citeauthoryear{{Bromm} \& {Yoshida}}{{Bromm} \& {Yoshida}}{2011}]{Bromm2011}
{Bromm} V.,  {Yoshida} N.,  2011, \mn@doi [\araa] {10.1146/annurev-astro-081710-102608}, \href {https://ui.adsabs.harvard.edu/abs/2011ARA&A..49..373B} {49, 373}

\bibitem[\protect\citeauthoryear{{Bromm}, {Yoshida}, {Hernquist}  \& {McKee}}{{Bromm} et~al.}{2009}]{Bromm2009}
{Bromm} V.,  {Yoshida} N.,  {Hernquist} L.,   {McKee} C.~F.,  2009, \mn@doi [\nat] {10.1038/nature07990}, \href {https://ui.adsabs.harvard.edu/abs/2009Natur.459...49B} {459, 49}

\bibitem[\protect\citeauthoryear{{Cardamone} et~al.,}{{Cardamone} et~al.}{2009}]{Cardamone2009}
{Cardamone} C.,  et~al., 2009, \mn@doi [\mnras] {10.1111/j.1365-2966.2009.15383.x}, \href {https://ui.adsabs.harvard.edu/abs/2009MNRAS.399.1191C} {399, 1191}

\bibitem[\protect\citeauthoryear{{Carrick}, {Turnbull}, {Lavaux}  \& {Hudson}}{{Carrick} et~al.}{2015}]{Carrick2015}
{Carrick} J.,  {Turnbull} S.~J.,  {Lavaux} G.,   {Hudson} M.~J.,  2015, \mn@doi [\mnras] {10.1093/mnras/stv547}, \href {https://ui.adsabs.harvard.edu/abs/2015MNRAS.450..317C} {450, 317}

\bibitem[\protect\citeauthoryear{{Chabrier}}{{Chabrier}}{2003}]{Chabrier2003}
{Chabrier} G.,  2003, \mn@doi [\pasp] {10.1086/376392}, \href {https://ui.adsabs.harvard.edu/abs/2003PASP..115..763C} {115, 763}

\bibitem[\protect\citeauthoryear{{Cheng} et~al.,}{{Cheng} et~al.}{2020}]{Cheng2020a}
{Cheng} T.-Y.,  et~al., 2020, \mn@doi [\mnras] {10.1093/mnras/staa501}, \href {https://ui.adsabs.harvard.edu/abs/2020MNRAS.493.4209C} {493, 4209}

\bibitem[\protect\citeauthoryear{{Cheng} et~al.,}{{Cheng} et~al.}{2021}]{Cheng2021b}
{Cheng} T.-Y.,  et~al., 2021, \mn@doi [\mnras] {10.1093/mnras/stab2142}, \href {https://ui.adsabs.harvard.edu/abs/2021MNRAS.507.4425C} {507, 4425}

\bibitem[\protect\citeauthoryear{{Cheng} et~al.,}{{Cheng} et~al.}{2023}]{Cheng2023}
{Cheng} T.~Y.,  et~al., 2023, \mn@doi [\mnras] {10.1093/mnras/stac3228}, \href {https://ui.adsabs.harvard.edu/abs/2023MNRAS.518.2794C} {518, 2794}

\bibitem[\protect\citeauthoryear{{Chiappini}, {Romano}  \& {Matteucci}}{{Chiappini} et~al.}{2003}]{Chiappini2003}
{Chiappini} C.,  {Romano} D.,   {Matteucci} F.,  2003, \mn@doi [\mnras] {10.1046/j.1365-8711.2003.06154.x}, \href {https://ui.adsabs.harvard.edu/abs/2003MNRAS.339...63C} {339, 63}

\bibitem[\protect\citeauthoryear{{Clemens}, {Crain}  \& {Anderson}}{{Clemens} et~al.}{2004}]{Clemens2004}
{Clemens} J.~C.,  {Crain} J.~A.,   {Anderson} R.,  2004, in {Moorwood} A. F.~M.,  {Iye} M.,  eds,  Society of Photo-Optical Instrumentation Engineers (SPIE) Conference Series Vol. 5492, Ground-based Instrumentation for Astronomy. pp 331--340, \mn@doi{10.1117/12.550069}

\bibitem[\protect\citeauthoryear{{Cooke}, {Pettini}, {Jorgenson}, {Murphy}  \& {Steidel}}{{Cooke} et~al.}{2014}]{Cooke2014}
{Cooke} R.~J.,  {Pettini} M.,  {Jorgenson} R.~A.,  {Murphy} M.~T.,   {Steidel} C.~C.,  2014, \mn@doi [\apj] {10.1088/0004-637X/781/1/31}, \href {https://ui.adsabs.harvard.edu/abs/2014ApJ...781...31C} {781, 31}

\bibitem[\protect\citeauthoryear{{Denicol{\'o}}, {Terlevich}  \& {Terlevich}}{{Denicol{\'o}} et~al.}{2002}]{Denicolo2002}
{Denicol{\'o}} G.,  {Terlevich} R.,   {Terlevich} E.,  2002, \mn@doi [\mnras] {10.1046/j.1365-8711.2002.05041.x}, \href {https://ui.adsabs.harvard.edu/abs/2002MNRAS.330...69D} {330, 69}

\bibitem[\protect\citeauthoryear{{Fern{\'a}ndez}, {Terlevich}, {D{\'\i}az}  \& {Terlevich}}{{Fern{\'a}ndez} et~al.}{2019}]{Fernandez2019}
{Fern{\'a}ndez} V.,  {Terlevich} E.,  {D{\'\i}az} A.~I.,   {Terlevich} R.,  2019, \mn@doi [\mnras] {10.1093/mnras/stz1433}, \href {https://ui.adsabs.harvard.edu/abs/2019MNRAS.487.3221F} {487, 3221}

\bibitem[\protect\citeauthoryear{{Frazier}}{{Frazier}}{2018}]{Frazier2018}
{Frazier} P.~I.,  2018, \mn@doi [arXiv e-prints] {10.48550/arXiv.1807.02811}, \href {https://ui.adsabs.harvard.edu/abs/2018arXiv180702811F} {p. arXiv:1807.02811}

\bibitem[\protect\citeauthoryear{{Fukugita} \& {Kawasaki}}{{Fukugita} \& {Kawasaki}}{2006}]{Fukugita2006}
{Fukugita} M.,  {Kawasaki} M.,  2006, \mn@doi [\apj] {10.1086/505109}, \href {https://ui.adsabs.harvard.edu/abs/2006ApJ...646..691F} {646, 691}

\bibitem[\protect\citeauthoryear{{Fukushima}, {Nagamine}, {Matsumoto}, {Isobe}, {Ouchi}, {Saitoh}  \& {Hirai}}{{Fukushima} et~al.}{2024}]{Fukushima2024}
{Fukushima} K.,  {Nagamine} K.,  {Matsumoto} A.,  {Isobe} Y.,  {Ouchi} M.,  {Saitoh} T.,   {Hirai} Y.,  2024, \mn@doi [arXiv e-prints] {10.48550/arXiv.2401.06450}, \href {https://ui.adsabs.harvard.edu/abs/2024arXiv240106450F} {p. arXiv:2401.06450}

\bibitem[\protect\citeauthoryear{{Goodfellow}, {Shlens}  \& {Szegedy}}{{Goodfellow} et~al.}{2014}]{Goodfellow2014}
{Goodfellow} I.~J.,  {Shlens} J.,   {Szegedy} C.,  2014, \mn@doi [arXiv e-prints] {10.48550/arXiv.1412.6572}, \href {https://ui.adsabs.harvard.edu/abs/2014arXiv1412.6572G} {p. arXiv:1412.6572}

\bibitem[\protect\citeauthoryear{{Griffith} et~al.,}{{Griffith} et~al.}{2011}]{Griffith2011}
{Griffith} R.~L.,  et~al., 2011, \mn@doi [\apjl] {10.1088/2041-8205/736/1/L22}, \href {https://ui.adsabs.harvard.edu/abs/2011ApJ...736L..22G} {736, L22}

\bibitem[\protect\citeauthoryear{{Grossi} et~al.,}{{Grossi} et~al.}{2025}]{Grossi2025}
{Grossi} M.,  et~al., 2025, arXiv e-prints, \href {https://ui.adsabs.harvard.edu/abs/2025arXiv250118498G} {p. arXiv:2501.18498}

\bibitem[\protect\citeauthoryear{{Groves}, {Heckman}  \& {Kauffmann}}{{Groves} et~al.}{2006}]{Groves2006}
{Groves} B.~A.,  {Heckman} T.~M.,   {Kauffmann} G.,  2006, \mn@doi [\mnras] {10.1111/j.1365-2966.2006.10812.x}, \href {https://ui.adsabs.harvard.edu/abs/2006MNRAS.371.1559G} {371, 1559}

\bibitem[\protect\citeauthoryear{{Guseva}, {Izotov}, {Papaderos}  \& {Fricke}}{{Guseva} et~al.}{2007}]{Guseva2007}
{Guseva} N.~G.,  {Izotov} Y.~I.,  {Papaderos} P.,   {Fricke} K.~J.,  2007, \mn@doi [\aap] {10.1051/0004-6361:20066067}, \href {https://ui.adsabs.harvard.edu/abs/2007A&A...464..885G} {464, 885}

\bibitem[\protect\citeauthoryear{{Guseva}, {Izotov}, {Fricke}  \& {Henkel}}{{Guseva} et~al.}{2017}]{Guseva2017}
{Guseva} N.~G.,  {Izotov} Y.~I.,  {Fricke} K.~J.,   {Henkel} C.,  2017, \mn@doi [\aap] {10.1051/0004-6361/201629181}, \href {https://ui.adsabs.harvard.edu/abs/2017A&A...599A..65G} {599, A65}

\bibitem[\protect\citeauthoryear{{Hirschauer} et~al.,}{{Hirschauer} et~al.}{2016}]{Hirschauer2016}
{Hirschauer} A.~S.,  et~al., 2016, \mn@doi [\apj] {10.3847/0004-637X/822/2/108}, \href {https://ui.adsabs.harvard.edu/abs/2016ApJ...822..108H} {822, 108}

\bibitem[\protect\citeauthoryear{{Hsyu}, {Cooke}, {Prochaska}  \& {Bolte}}{{Hsyu} et~al.}{2017}]{Hsyu2017}
{Hsyu} T.,  {Cooke} R.~J.,  {Prochaska} J.~X.,   {Bolte} M.,  2017, \mn@doi [\apjl] {10.3847/2041-8213/aa821f}, \href {https://ui.adsabs.harvard.edu/abs/2017ApJ...845L..22H} {845, L22}

\bibitem[\protect\citeauthoryear{{Hsyu}, {Cooke}, {Prochaska}  \& {Bolte}}{{Hsyu} et~al.}{2018}]{Hsyu2018}
{Hsyu} T.,  {Cooke} R.~J.,  {Prochaska} J.~X.,   {Bolte} M.,  2018, \mn@doi [\apj] {10.3847/1538-4357/aad18a}, \href {https://ui.adsabs.harvard.edu/abs/2018ApJ...863..134H} {863, 134}

\bibitem[\protect\citeauthoryear{{Hsyu}, {Cooke}, {Prochaska}  \& {Bolte}}{{Hsyu} et~al.}{2020}]{Hsyu2020}
{Hsyu} T.,  {Cooke} R.~J.,  {Prochaska} J.~X.,   {Bolte} M.,  2020, \mn@doi [\apj] {10.3847/1538-4357/ab91af}, \href {https://ui.adsabs.harvard.edu/abs/2020ApJ...896...77H} {896, 77}

\bibitem[\protect\citeauthoryear{{Isobe} et~al.,}{{Isobe} et~al.}{2022}]{Isobe2022empress}
{Isobe} Y.,  et~al., 2022, \mn@doi [\apj] {10.3847/1538-4357/ac3509}, \href {https://ui.adsabs.harvard.edu/abs/2022ApJ...925..111I} {925, 111}

\bibitem[\protect\citeauthoryear{{Izotov} \& {Thuan}}{{Izotov} \& {Thuan}}{2007}]{Izotov2007}
{Izotov} Y.~I.,  {Thuan} T.~X.,  2007, \mn@doi [\apj] {10.1086/519922}, \href {https://ui.adsabs.harvard.edu/abs/2007ApJ...665.1115I} {665, 1115}

\bibitem[\protect\citeauthoryear{{Izotov} \& {Thuan}}{{Izotov} \& {Thuan}}{2009}]{IzotovThuan2009}
{Izotov} Y.~I.,  {Thuan} T.~X.,  2009, \mn@doi [\apj] {10.1088/0004-637X/690/2/1797}, \href {https://ui.adsabs.harvard.edu/abs/2009ApJ...690.1797I} {690, 1797}

\bibitem[\protect\citeauthoryear{{Izotov}, {Lipovetsky}, {Chaffee}, {Foltz}, {Guseva}  \& {Kniazev}}{{Izotov} et~al.}{1997}]{Izotov1997}
{Izotov} Y.~I.,  {Lipovetsky} V.~A.,  {Chaffee} F.~H.,  {Foltz} C.~B.,  {Guseva} N.~G.,   {Kniazev} A.~Y.,  1997, \mn@doi [\apj] {10.1086/303664}, \href {https://ui.adsabs.harvard.edu/abs/1997ApJ...476..698I} {476, 698}

\bibitem[\protect\citeauthoryear{{Izotov}, {Papaderos}, {Guseva}, {Fricke}  \& {Thuan}}{{Izotov} et~al.}{2006}]{Izotov2006}
{Izotov} Y.~I.,  {Papaderos} P.,  {Guseva} N.~G.,  {Fricke} K.~J.,   {Thuan} T.~X.,  2006, \mn@doi [\aap] {10.1051/0004-6361:20065100}, \href {https://ui.adsabs.harvard.edu/abs/2006A&A...454..137I} {454, 137}

\bibitem[\protect\citeauthoryear{{Izotov}, {Guseva}, {Fricke}  \& {Papaderos}}{{Izotov} et~al.}{2009}]{Izotov2009}
{Izotov} Y.~I.,  {Guseva} N.~G.,  {Fricke} K.~J.,   {Papaderos} P.,  2009, \mn@doi [\aap] {10.1051/0004-6361/200911965}, \href {https://ui.adsabs.harvard.edu/abs/2009A&A...503...61I} {503, 61}

\bibitem[\protect\citeauthoryear{{Izotov}, {Thuan}  \& {Guseva}}{{Izotov} et~al.}{2012}]{Izotov2012}
{Izotov} Y.~I.,  {Thuan} T.~X.,   {Guseva} N.~G.,  2012, \mn@doi [\aap] {10.1051/0004-6361/201219733}, \href {https://ui.adsabs.harvard.edu/abs/2012A&A...546A.122I} {546, A122}

\bibitem[\protect\citeauthoryear{{Izotov}, {Thuan}  \& {Guseva}}{{Izotov} et~al.}{2014}]{Izotov2014}
{Izotov} Y.~I.,  {Thuan} T.~X.,   {Guseva} N.~G.,  2014, \mn@doi [\mnras] {10.1093/mnras/stu1771}, \href {https://ui.adsabs.harvard.edu/abs/2014MNRAS.445..778I} {445, 778}

\bibitem[\protect\citeauthoryear{{James}, {Koposov}, {Stark}, {Belokurov}, {Pettini}, {Olszewski}  \& {McQuinn}}{{James} et~al.}{2017}]{James2017}
{James} B.~L.,  {Koposov} S.~E.,  {Stark} D.~P.,  {Belokurov} V.,  {Pettini} M.,  {Olszewski} E.~W.,   {McQuinn} K. B.~W.,  2017, \mn@doi [\mnras] {10.1093/mnras/stw2962}, \href {https://ui.adsabs.harvard.edu/abs/2017MNRAS.465.3977J} {465, 3977}

\bibitem[\protect\citeauthoryear{{Karachentsev}, {Makarova}, {Koribalski}, {Anand}, {Tully}  \& {Kniazev}}{{Karachentsev} et~al.}{2023}]{Karachentsev2023}
{Karachentsev} I.~D.,  {Makarova} L.~N.,  {Koribalski} B.~S.,  {Anand} G.~S.,  {Tully} R.~B.,   {Kniazev} A.~Y.,  2023, \mn@doi [\mnras] {10.1093/mnras/stac3284}, \href {https://ui.adsabs.harvard.edu/abs/2023MNRAS.518.5893K} {518, 5893}

\bibitem[\protect\citeauthoryear{{Kauffmann} et~al.,}{{Kauffmann} et~al.}{2003a}]{Kauffmann2003a}
{Kauffmann} G.,  et~al., 2003a, \mn@doi [\mnras] {10.1046/j.1365-8711.2003.06291.x}, \href {https://ui.adsabs.harvard.edu/abs/2003MNRAS.341...33K} {341, 33}

\bibitem[\protect\citeauthoryear{{Kauffmann} et~al.,}{{Kauffmann} et~al.}{2003b}]{Kauffmann2003b}
{Kauffmann} G.,  et~al., 2003b, \mn@doi [\mnras] {10.1111/j.1365-2966.2003.07154.x}, \href {https://ui.adsabs.harvard.edu/abs/2003MNRAS.346.1055K} {346, 1055}

\bibitem[\protect\citeauthoryear{{Kennicutt}}{{Kennicutt}}{1998}]{Kennicutt1998}
{Kennicutt} Robert~C. J.,  1998, \mn@doi [\araa] {10.1146/annurev.astro.36.1.189}, \href {https://ui.adsabs.harvard.edu/abs/1998ARA&A..36..189K} {36, 189}

\bibitem[\protect\citeauthoryear{{Kennicutt}, {Tamblyn}  \& {Congdon}}{{Kennicutt} et~al.}{1994}]{Kennicutt1994}
{Kennicutt} Robert~C. J.,  {Tamblyn} P.,   {Congdon} C.~E.,  1994, \mn@doi [\apj] {10.1086/174790}, \href {https://ui.adsabs.harvard.edu/abs/1994ApJ...435...22K} {435, 22}

\bibitem[\protect\citeauthoryear{{Kewley}, {Groves}, {Kauffmann}  \& {Heckman}}{{Kewley} et~al.}{2006}]{Kewley2006}
{Kewley} L.~J.,  {Groves} B.,  {Kauffmann} G.,   {Heckman} T.,  2006, \mn@doi [\mnras] {10.1111/j.1365-2966.2006.10859.x}, \href {https://ui.adsabs.harvard.edu/abs/2006MNRAS.372..961K} {372, 961}

\bibitem[\protect\citeauthoryear{{Kewley}, {Dopita}, {Leitherer}, {Dav{\'e}}, {Yuan}, {Allen}, {Groves}  \& {Sutherland}}{{Kewley} et~al.}{2013}]{Kewley2013}
{Kewley} L.~J.,  {Dopita} M.~A.,  {Leitherer} C.,  {Dav{\'e}} R.,  {Yuan} T.,  {Allen} M.,  {Groves} B.,   {Sutherland} R.,  2013, \mn@doi [\apj] {10.1088/0004-637X/774/2/100}, \href {https://ui.adsabs.harvard.edu/abs/2013ApJ...774..100K} {774, 100}

\bibitem[\protect\citeauthoryear{Kingma \& Ba}{Kingma \& Ba}{2015}]{Kingma2015}
Kingma D.~P.,  Ba J.,  2015, in Bengio Y.,  LeCun Y.,  eds, 3rd International Conference on Learning Representations, {ICLR} 2015, San Diego, CA, USA, May 7-9, 2015, Conference Track Proceedings. \url {http://arxiv.org/abs/1412.6980}

\bibitem[\protect\citeauthoryear{{Kojima} et~al.,}{{Kojima} et~al.}{2020}]{Kojima2020empress}
{Kojima} T.,  et~al., 2020, \mn@doi [\apj] {10.3847/1538-4357/aba047}, \href {https://ui.adsabs.harvard.edu/abs/2020ApJ...898..142K} {898, 142}

\bibitem[\protect\citeauthoryear{{Kunth} \& {{\"O}stlin}}{{Kunth} \& {{\"O}stlin}}{2000}]{Kunth2000}
{Kunth} D.,  {{\"O}stlin} G.,  2000, \mn@doi [\aapr] {10.1007/s001590000005}, \href {https://ui.adsabs.harvard.edu/abs/2000A&ARv..10....1K} {10, 1}

\bibitem[\protect\citeauthoryear{{Liu}, {Luo}, {Zhang}, {Kong}, {Zhang}, {Shen}  \& {Zhao}}{{Liu} et~al.}{2023}]{Liu2023}
{Liu} S.,  {Luo} A.~L.,  {Zhang} W.,  {Kong} X.,  {Zhang} Y.-X.,  {Shen} S.-Y.,   {Zhao} Y.-H.,  2023, \mn@doi [\apjs] {10.3847/1538-4365/acd69c}, \href {https://ui.adsabs.harvard.edu/abs/2023ApJS..267...16L} {267, 16}

\bibitem[\protect\citeauthoryear{{Madau}, {Pozzetti}  \& {Dickinson}}{{Madau} et~al.}{1998}]{Madau1998}
{Madau} P.,  {Pozzetti} L.,   {Dickinson} M.,  1998, \mn@doi [\apj] {10.1086/305523}, \href {https://ui.adsabs.harvard.edu/abs/1998ApJ...498..106M} {498, 106}

\bibitem[\protect\citeauthoryear{{Marigo}}{{Marigo}}{2001}]{Marigo2001}
{Marigo} P.,  2001, \mn@doi [\aap] {10.1051/0004-6361:20000247}, \href {https://ui.adsabs.harvard.edu/abs/2001A&A...370..194M} {370, 194}

\bibitem[\protect\citeauthoryear{{Matsumoto} et~al.,}{{Matsumoto} et~al.}{2022}]{Matsumoto2022}
{Matsumoto} A.,  et~al., 2022, \mn@doi [\apj] {10.3847/1538-4357/ac9ea1}, \href {https://ui.adsabs.harvard.edu/abs/2022ApJ...941..167M} {941, 167}

\bibitem[\protect\citeauthoryear{{McQuinn} et~al.,}{{McQuinn} et~al.}{2015}]{McQuinn2015}
{McQuinn} K. B.~W.,  et~al., 2015, \mn@doi [\apj] {10.1088/0004-637X/812/2/158}, \href {https://ui.adsabs.harvard.edu/abs/2015ApJ...812..158M} {812, 158}

\bibitem[\protect\citeauthoryear{{Meynet} \& {Maeder}}{{Meynet} \& {Maeder}}{2002}]{Meynet2002}
{Meynet} G.,  {Maeder} A.,  2002, \mn@doi [\aap] {10.1051/0004-6361:20020755}, \href {https://ui.adsabs.harvard.edu/abs/2002A&A...390..561M} {390, 561}

\bibitem[\protect\citeauthoryear{{Micheva}, {{\"O}stlin}, {Bergvall}, {Zackrisson}, {Masegosa}, {Marquez}, {Marquart}  \& {Durret}}{{Micheva} et~al.}{2013}]{Micheva2013}
{Micheva} G.,  {{\"O}stlin} G.,  {Bergvall} N.,  {Zackrisson} E.,  {Masegosa} J.,  {Marquez} I.,  {Marquart} T.,   {Durret} F.,  2013, \mn@doi [\mnras] {10.1093/mnras/stt146}, \href {https://ui.adsabs.harvard.edu/abs/2013MNRAS.431..102M} {431, 102}

\bibitem[\protect\citeauthoryear{{Nakajima} et~al.,}{{Nakajima} et~al.}{2022}]{Nakajima2022empress}
{Nakajima} K.,  et~al., 2022, \mn@doi [\apjs] {10.3847/1538-4365/ac7710}, \href {https://ui.adsabs.harvard.edu/abs/2022ApJS..262....3N} {262, 3}

\bibitem[\protect\citeauthoryear{{Peimbert}, {Peimbert}  \& {Luridiana}}{{Peimbert} et~al.}{2016}]{Peimbert2016}
{Peimbert} A.,  {Peimbert} M.,   {Luridiana} V.,  2016, \mn@doi [\rmxaa] {10.48550/arXiv.1608.02062}, \href {https://ui.adsabs.harvard.edu/abs/2016RMxAA..52..419P} {52, 419}

\bibitem[\protect\citeauthoryear{{Pettini} \& {Pagel}}{{Pettini} \& {Pagel}}{2004}]{Pettini2004}
{Pettini} M.,  {Pagel} B. E.~J.,  2004, \mn@doi [\mnras] {10.1111/j.1365-2966.2004.07591.x}, \href {https://ui.adsabs.harvard.edu/abs/2004MNRAS.348L..59P} {348, L59}

\bibitem[\protect\citeauthoryear{{Pilyugin}, {Thuan}  \& {V{\'\i}lchez}}{{Pilyugin} et~al.}{2003}]{Pilyugin2003}
{Pilyugin} L.~S.,  {Thuan} T.~X.,   {V{\'\i}lchez} J.~M.,  2003, \mn@doi [\aap] {10.1051/0004-6361:20021458}, \href {https://ui.adsabs.harvard.edu/abs/2003A&A...397..487P} {397, 487}

\bibitem[\protect\citeauthoryear{{Planck Collaboration} et~al.,}{{Planck Collaboration} et~al.}{2020}]{Planck2020}
{Planck Collaboration} et~al., 2020, \mn@doi [\aap] {10.1051/0004-6361/201833910}, \href {https://ui.adsabs.harvard.edu/abs/2020A&A...641A...6P} {641, A6}

\bibitem[\protect\citeauthoryear{{Prochaska} et~al.,}{{Prochaska} et~al.}{2020a}]{pypeit:zenodo}
{Prochaska} J.~X.,  et~al., 2020a, {pypeit/PypeIt: Release 1.0.0}, \mn@doi{10.5281/zenodo.3743493}

\bibitem[\protect\citeauthoryear{Prochaska et~al.,}{Prochaska et~al.}{2020b}]{pypeit:joss_pub}
Prochaska J.~X.,  et~al., 2020b, \mn@doi [Journal of Open Source Software] {10.21105/joss.02308}, 5, 2308

\bibitem[\protect\citeauthoryear{{Pustilnik}, {Tepliakova}, {Kniazev}, {Martin}  \& {Burenkov}}{{Pustilnik} et~al.}{2010}]{Pustilnik2010}
{Pustilnik} S.~A.,  {Tepliakova} A.~L.,  {Kniazev} A.~Y.,  {Martin} J.~M.,   {Burenkov} A.~N.,  2010, \mn@doi [\mnras] {10.1111/j.1365-2966.2009.15637.x}, \href {https://ui.adsabs.harvard.edu/abs/2010MNRAS.401..333P} {401, 333}

\bibitem[\protect\citeauthoryear{{Raimann}, {Bica}, {Storchi-Bergmann}, {Melnick}  \& {Schmitt}}{{Raimann} et~al.}{2000}]{Raimann2000}
{Raimann} D.,  {Bica} E.,  {Storchi-Bergmann} T.,  {Melnick} J.,   {Schmitt} H.,  2000, \mn@doi [\mnras] {10.1046/j.1365-8711.2000.03317.x}, \href {https://ui.adsabs.harvard.edu/abs/2000MNRAS.314..295R} {314, 295}

\bibitem[\protect\citeauthoryear{{Roy}, {Dopita}, {Krumholz}, {Kewley}, {Sutherland}  \& {Heger}}{{Roy} et~al.}{2021}]{Roy2021}
{Roy} A.,  {Dopita} M.~A.,  {Krumholz} M.~R.,  {Kewley} L.~J.,  {Sutherland} R.~S.,   {Heger} A.,  2021, \mn@doi [\mnras] {10.1093/mnras/stab376}, \href {https://ui.adsabs.harvard.edu/abs/2021MNRAS.502.4359R} {502, 4359}

\bibitem[\protect\citeauthoryear{{Ruiz-Escobedo}, {Pe{\~n}a}, {Hern{\'a}ndez-Mart{\'\i}nez}  \& {Garc{\'\i}a-Rojas}}{{Ruiz-Escobedo} et~al.}{2018}]{Ruiz-Escobedo2018}
{Ruiz-Escobedo} F.,  {Pe{\~n}a} M.,  {Hern{\'a}ndez-Mart{\'\i}nez} L.,   {Garc{\'\i}a-Rojas} J.,  2018, \mn@doi [\mnras] {10.1093/mnras/sty2265}, \href {https://ui.adsabs.harvard.edu/abs/2018MNRAS.481..396R} {481, 396}

\bibitem[\protect\citeauthoryear{{Salpeter}}{{Salpeter}}{1955}]{Salpeter1955}
{Salpeter} E.~E.,  1955, \mn@doi [\apj] {10.1086/145971}, \href {https://ui.adsabs.harvard.edu/abs/1955ApJ...121..161S} {121, 161}

\bibitem[\protect\citeauthoryear{{Sargent} \& {Searle}}{{Sargent} \& {Searle}}{1970}]{Sargent1970}
{Sargent} W. L.~W.,  {Searle} L.,  1970, \mn@doi [\apjl] {10.1086/180644}, \href {https://ui.adsabs.harvard.edu/abs/1970ApJ...162L.155S} {162, L155}

\bibitem[\protect\citeauthoryear{{Skillman} et~al.,}{{Skillman} et~al.}{2013}]{Skillman2013}
{Skillman} E.~D.,  et~al., 2013, \mn@doi [\aj] {10.1088/0004-6256/146/1/3}, \href {https://ui.adsabs.harvard.edu/abs/2013AJ....146....3S} {146, 3}

\bibitem[\protect\citeauthoryear{{Steigman}}{{Steigman}}{2007}]{Steigman2007}
{Steigman} G.,  2007, \mn@doi [Annual Review of Nuclear and Particle Science] {10.1146/annurev.nucl.56.080805.140437}, \href {https://ui.adsabs.harvard.edu/abs/2007ARNPS..57..463S} {57, 463}

\bibitem[\protect\citeauthoryear{{Storchi-Bergmann}, {Calzetti}  \& {Kinney}}{{Storchi-Bergmann} et~al.}{1994}]{Storchi-Bergmann1994}
{Storchi-Bergmann} T.,  {Calzetti} D.,   {Kinney} A.~L.,  1994, \mn@doi [\apj] {10.1086/174345}, \href {https://ui.adsabs.harvard.edu/abs/1994ApJ...429..572S} {429, 572}

\bibitem[\protect\citeauthoryear{{Tachiev} \& {Froese Fischer}}{{Tachiev} \& {Froese Fischer}}{2001}]{Tachiev2001}
{Tachiev} G.,  {Froese Fischer} C.,  2001, \mn@doi [Canadian Journal of Physics] {10.1139/p01-059}, \href {https://ui.adsabs.harvard.edu/abs/2001CaJPh..79..955T} {79, 955}

\bibitem[\protect\citeauthoryear{{Thuan} \& {Izotov}}{{Thuan} \& {Izotov}}{2005}]{Thuan2005}
{Thuan} T.~X.,  {Izotov} Y.~I.,  2005, \mn@doi [\apjs] {10.1086/491657}, \href {https://ui.adsabs.harvard.edu/abs/2005ApJS..161..240T} {161, 240}

\bibitem[\protect\citeauthoryear{{Thuan}, {Izotov}  \& {Lipovetsky}}{{Thuan} et~al.}{1995}]{Thuan1995}
{Thuan} T.~X.,  {Izotov} Y.~I.,   {Lipovetsky} V.~A.,  1995, \mn@doi [\apj] {10.1086/175676}, \href {https://ui.adsabs.harvard.edu/abs/1995ApJ...445..108T} {445, 108}

\bibitem[\protect\citeauthoryear{{Thuan}, {Guseva}  \& {Izotov}}{{Thuan} et~al.}{2022}]{Thuan2022}
{Thuan} T.~X.,  {Guseva} N.~G.,   {Izotov} Y.~I.,  2022, \mn@doi [\mnras] {10.1093/mnrasl/slac095}, \href {https://ui.adsabs.harvard.edu/abs/2022MNRAS.516L..81T} {516, L81}

\bibitem[\protect\citeauthoryear{{Tremonti} et~al.,}{{Tremonti} et~al.}{2004}]{Tremonti2004}
{Tremonti} C.~A.,  et~al., 2004, \mn@doi [\apj] {10.1086/423264}, \href {https://ui.adsabs.harvard.edu/abs/2004ApJ...613..898T} {613, 898}

\bibitem[\protect\citeauthoryear{{Wang} et~al.,}{{Wang} et~al.}{2018}]{Wang2018}
{Wang} L.-L.,  et~al., 2018, \mn@doi [\mnras] {10.1093/mnras/stx2798}, \href {https://ui.adsabs.harvard.edu/abs/2018MNRAS.474.1873W} {474, 1873}

\bibitem[\protect\citeauthoryear{{Wise}, {Turk}, {Norman}  \& {Abel}}{{Wise} et~al.}{2012}]{Wise2012}
{Wise} J.~H.,  {Turk} M.~J.,  {Norman} M.~L.,   {Abel} T.,  2012, \mn@doi [\apj] {10.1088/0004-637X/745/1/50}, \href {https://ui.adsabs.harvard.edu/abs/2012ApJ...745...50W} {745, 50}

\bibitem[\protect\citeauthoryear{{Xu} et~al.,}{{Xu} et~al.}{2022}]{Xu2022ApJempress}
{Xu} Y.,  et~al., 2022, \mn@doi [\apj] {10.3847/1538-4357/ac5e32}, \href {https://ui.adsabs.harvard.edu/abs/2022ApJ...929..134X} {929, 134}

\bibitem[\protect\citeauthoryear{{Yang}, {Malhotra}, {Rhoads}  \& {Wang}}{{Yang} et~al.}{2017}]{Yang2017}
{Yang} H.,  {Malhotra} S.,  {Rhoads} J.~E.,   {Wang} J.,  2017, \mn@doi [\apj] {10.3847/1538-4357/aa8809}, \href {https://ui.adsabs.harvard.edu/abs/2017ApJ...847...38Y} {847, 38}

\bibitem[\protect\citeauthoryear{{Yin}, {Liang}, {Hammer}, {Brinchmann}, {Zhang}, {Deng}  \& {Flores}}{{Yin} et~al.}{2007}]{Yin2007}
{Yin} S.~Y.,  {Liang} Y.~C.,  {Hammer} F.,  {Brinchmann} J.,  {Zhang} B.,  {Deng} L.~C.,   {Flores} H.,  2007, \mn@doi [\aap] {10.1051/0004-6361:20065798}, \href {https://ui.adsabs.harvard.edu/abs/2007A&A...462..535Y} {462, 535}

\bibitem[\protect\citeauthoryear{{Zibetti}, {Charlot}  \& {Rix}}{{Zibetti} et~al.}{2009}]{Zibetti2009}
{Zibetti} S.,  {Charlot} S.,   {Rix} H.-W.,  2009, \mn@doi [\mnras] {10.1111/j.1365-2966.2009.15528.x}, \href {https://ui.adsabs.harvard.edu/abs/2009MNRAS.400.1181Z} {400, 1181}

\bibitem[\protect\citeauthoryear{{Zou} et~al.,}{{Zou} et~al.}{2024}]{Zou2024}
{Zou} H.,  et~al., 2024, \mn@doi [\apj] {10.3847/1538-4357/ad1409}, \href {https://ui.adsabs.harvard.edu/abs/2024ApJ...961..173Z} {961, 173}

\bibitem[\protect\citeauthoryear{{van Zee}}{{van Zee}}{2000}]{vanZee2000}
{van Zee} L.,  2000, \mn@doi [\apjl] {10.1086/318176}, \href {https://ui.adsabs.harvard.edu/abs/2000ApJ...543L..31V} {543, L31}

\bibitem[\protect\citeauthoryear{{van Zee} \& {Haynes}}{{van Zee} \& {Haynes}}{2006}]{vanZee2006}
{van Zee} L.,  {Haynes} M.~P.,  2006, \mn@doi [\apj] {10.1086/498017}, \href {https://ui.adsabs.harvard.edu/abs/2006ApJ...636..214V} {636, 214}

\makeatother
\end{thebibliography}

\appendix


\label{lastpage}
\end{document}